%% file: main.tex
\documentclass{article}

\usepackage[eandd, final]{neurips_2026}

\usepackage[utf8]{inputenc} 
\usepackage[T1]{fontenc}    
\usepackage{hyperref}       
\usepackage{url}            
\usepackage{booktabs}       
\usepackage{amsfonts}       
\usepackage{nicefrac}       
\usepackage{microtype}      
\usepackage{xcolor} 
\usepackage{colortbl}
\usepackage{adjustbox}
\usepackage{wrapfig}
\usepackage{longtable}
\usepackage{subcaption}
\usepackage{amsmath}
\usepackage{float}
\usepackage{multirow}
\usepackage{graphicx}
\usepackage{inconsolata}
\usepackage{xspace}
\usepackage{placeins}
\usepackage{tabularx}
\usepackage{array}
\newcolumntype{Y}{>{\raggedright\arraybackslash}X}

\usepackage{ulem}
\newcommand{\bench}{\textit{CellularS2-Bench}\xspace}
\newcommand{\ari}{\textit{CellularS2-ARI}\xspace}

\definecolor{stage1color}{HTML}{CFE2F3}
\definecolor{stage2color}{HTML}{FCE5CD}
\definecolor{stage3color}{HTML}{D9D2E9}

\title{
\bench: A Staged, Evidence-Grounded Benchmark for Cellular Network Security Reasoning
}

\author{%
  \resizebox{0.98\textwidth}{!}{\mbox{%
    Ke Xie\textsuperscript{1}\quad
    Xingyi Zhao\textsuperscript{1}\quad
    Yiwen Hu\textsuperscript{2}\quad
    Haotian Lin\textsuperscript{3}\thanks{This work does not relate to the author's position at Amazon.}\quad
    Shuhan Yuan\textsuperscript{1}\quad
    Wen Li\textsuperscript{1}\quad
    Tian Xie\textsuperscript{1}%
  }}\\[0.6em]
  \normalfont\small
  \textsuperscript{1}Utah State University\qquad
  \textsuperscript{2}University of Maryland, Baltimore County\qquad
  \textsuperscript{3}Amazon\\[0.4em]
}
\begin{document}

\maketitle

\begin{abstract}
Cellular networks are critical infrastructure supporting billions of worldwide users and safety- and mission-critical services. Vulnerabilities in cellular networks can therefore cause service disruption, privacy breaches, and broad societal harm, motivating growing efforts to analyze 3GPP specifications that define required device and operator behavior.  However, existing benchmarks for this domain share three systematic deficiencies: they rely on LLM-generated entries without expert verification, they do not require faithful interpretation of normative language, reasoning across cross-referenced clauses, and verifiable conclusions grounded in multimodal evidence such as tables and figures, and they do not evaluate security reasoning over end-to-end protocol behavior. To address these gaps, we present \bench, a staged benchmark. \bench covers core control-plane specifications, processing over 
1,866 paragraphs, 22,033 sentences, 424 figures, and 248 tables. 
It comprises 13,851 questions and instances across three progressive stages, of which 10,269 (74.14\%) are newly constructed from 3GPP specifications and 43 real-world vulnerabilities. Beyond newly constructed tasks, \bench integrates and corrects 3,582 (25.86\%)
instances from 7 prior datasets to broaden task coverage. To establish a reproducible baseline on \bench, we introduce \ari, a unified Adapt--Retrieve--Integrate framework for systematic understanding and specification-driven security analysis of 3GPP specifications. Together, they establish an accessible and reproducible foundation for quantifying progress in specification understanding and security reasoning in cellular network security.

\end{abstract}

\input{latex/intro}

\input{latex/background}
\input{latex/related_works}

\input{latex/challenge}
\input{latex/benchmark}

\input{latex/methods}

\input{latex/evaluation}

\input{latex/limitation}

\input{latex/conclusion}

\newpage
\bibliographystyle{plainnat}
\bibliography{latex/custom}

\input{latex/appendix}

\input{checklist.tex}

\end{document}

%% file: latex/intro.tex
\section{Introduction}

Cellular networks have become the critical infrastructure of modern society. 
They serve not only billions of users worldwide, but also safety- and mission-critical services such as financial transactions, public safety systems, and medical communications. Therefore, security vulnerabilities in cellular networks can have severe consequences: once exploited, they may lead to service disruptions, privacy breaches, and large-scale societal harm. Ensuring the security of mobile networks has thus long been a fundamental and enduring challenge. 
Beyond code review~\cite{10.1145/3745019} and fuzz testing of network implementations and functions~\cite{godefroid2020fuzzing}, 
increasing attention has been devoted to analyzing 3GPP specifications, which define required behavior of devices and operators worldwide.  

However, analyzing 3GPP specifications is challenging. These standards use highly domain-specific language and interleave dense normative statements with tables and figures. Crucial details are often scattered across cross-referenced clauses and even multiple specifications. Correct security analysis requires more than locating relevant text: it requires domain-specific multi-source, cross-clause integration; precise interpretation of multimodal context; and evidence-grounded reasoning over procedures and security properties.

Existing efforts on cellular-standard security analysis largely rely on conventional NLP techniques, such as sentence-pair semantic relation identification (e.g., entailment, conflict)~\cite{287360,298168}. Recent LLM-based frameworks~\cite{maatouk2025telellmsseriesspecializedlarge, nikbakht2024tspecllmopensourcedatasetllm,10.1145/3795866.3844474} have begun to explore pilot architectures for interpreting 3GPP specifications.  

\textit{However, while prior studies have identified and analyzed some cellular vulnerabilities, the community lacks a comprehensive and reproducible benchmark for evaluating 3GPP specification understanding and security reasoning. As a result, it remains difficult to quantify progress or compare systems on the safety-critical reasoning capabilities required by both research and industry communities.}

To address the gap, we present \bench (\underline{Cellular} \underline{S}pecification \underline{S}ecurity \underline{Bench}mark) and \ari (\underline{Cellular} \underline{S}pecification \underline{S}ecurity Analysis based on \underline{A}dapt-\underline{R}etrieve--\underline{I}ntegrate). \bench adopts a three-stage design that directly mirrors the three capability layers required for rigorous spec interpretation: Stage 1 targets intra-clause understanding; Stage 2 adds evidence-grounded retrieval and citation; Stage 3 requires cross-clause integration and security-aware reasoning. It covers core Release-17 specifications, including specifications in the TS~24 series and
TS~38.331, which define essential cellular procedures such as registration,
authentication, mobility management, and radio signaling, processing over 
1,866 paragraphs, 22,033 sentences, 424 figures, and 248 tables,
capturing both textual and multimodal specification semantics. In total, \bench comprises 13,851 
instances. Of these, 10,269 (74.14\%) are newly derived from 3GPP specifications and 43 real-world vulnerabilities reported in prior works~\cite{shaik2015practical, 6550445, kambourakis2011attacks, lee2009detection, leong2014unveiling, kim2019touching, van2015defeating, park2016white, chlosta2019lte, 8958725, michau2016not, hussain20195greasoner, borgaonkar2018new, 8894379, chlosta20215g, hussain2019privacy, al2024hermes, hussain2018lteinspector, 10.1145/3795866.3844474}. 
In addition to newly constructed questions, \bench selectively adopts 3,582 instances (25.86\%) from prior datasets
to broaden its coverage of question types and reasoning requirements. To validate that \bench poses meaningful challenges and to provide reproducible baselines, we design \ari, a unified framework to solve tasks in (1) interpreting complex 3GPP specifications written in domain-specific language with multimodal context, and (2) performing the challenging security analysis over cellular network specifications.

In summary, we make three key contributions:  (1) we establish \bench, a staged benchmark with 10,269 newly constructed instances to evaluate frameworks' capabilities of 3GPP specifications interpretation and security reasoning, including expert-constructed security tasks grounded in 43 real-world vulnerabilities; (2) \bench selectively integrates 3,582 compatible instances from prior datasets after careful correctness validation, improving coverage while maintaining reliability; (3) we present a unified framework, \ari, for systematic analysis of 3GPP specifications, and apply it to \bench to provide baseline results.

%% file: latex/background.tex
\section{Background - 3GPP Specifications and Threats}

The 3rd Generation Partnership Project (3GPP), an association of seven Organizational Partners, including ATIS in the U.S., ETSI in Europe, and CCSA in China, standardized the cellular network architectures and services via technical specifications (i.e., 3GPP specifications) for the operational 4G/5G networks to ongoing 6G network. Thousands of specifications are published by 3GPP to ensure global interoperability across mobile networks and devices. Appendix~\ref{app:glossary} provides a glossary of cellular network terminology.
Mobile equipment vendors and operators are required to ensure compliance with 3GPP specifications. However, this standardization introduces systemic risks: design flaws in 3GPP specifications can propagate into mobile networks on a global scale. 
Moreover, the increasing complexity, frequent revisions, and evolving new generations make 3GPP specifications one of the most comprehensive and technically intricate corpora in the cellular network domain. 

%% file: latex/related_works.tex
\section{Related Work}

\noindent \textbf{Cellular Network QA and Retrieval Benchmarks.}
Recent benchmarks evaluate LLM-based QA (question answering) and retrieval over telecom and 3GPP resources. TSpec-LLM~\cite{nikbakht2024tspecllmopensourcedatasetllm} releases a large-scale corpus of 3GPP specifications (Release 8--19). Telco-DPR~\cite{saraiva2024telcodprhybriddatasetevaluating} targets RAG retrieval through synthetic QA pairs. TeleQnA~\cite{maatouk2023teleqnabenchmarkdatasetassess} provides 10K multiple-choice questions from specifications and broader sources, including research literature, and a telecom lexicon. 
TeleQuAD~\cite{telequad2025} provides a suite of telecom question answering tasks, including extractive QA over specification text as well as tabular question answering over specification tables.
Tele-LLMs~\cite{maatouk2025telellmsseriesspecializedlarge} builds Tele-Data from specifications and an open-ended QA set for domain knowledge adaptation evaluation. 
Despite broad coverage, they primarily evaluate the understanding of a single document rather than task-driven reasoning, such as integrating specification evidence to identify security vulnerabilities. Such task-driven reasoning is closer to how experts analyze standards in practice. 
Moreover, these datasets rely on LLM-generated entries without expert verification, which limits their reliability for safety-critical evaluation. In contrast, \bench emphasizes expert-verified, evidence-grounded, and security-oriented reasoning over 3GPP specifications, including tasks that require cross-clause integration and multimodal evidence from text, tables, and figures.

\noindent \textbf{Cellular Network Specific Task Datasets.} 
Several datasets introduce telecom-oriented tasks over specifications. SPEC5G~\cite{karim2023spec5gdataset5gcellular} builds a 5G-focused corpus from specifications, augmented with content scraped from blogs and technical forums, and defines two tasks: security-related text classification and specification summarization. 
CellularLint~\cite{298168} and ConTester~\cite{287360} derive sentence pairs from 3GPP specifications and annotate fine-grained semantic relations to support Natural Language Inference (NLI) reasoning and inconsistency detection. 
While valuable, these datasets have notable limitations: (1) they primarily measure sentence-/paragraph-level understanding (classification, summarization, or pairwise relations) rather than standards-driven procedural reasoning that requires integrating evidence across multiple clauses/tables/figures; (2) they do not explicitly evaluate evidence-grounded performance, which is critical for complex reasoning that conclusions must be traceable; (3) their tasks do not explicitly measure the security reasoning capability. Classifying security-relevant text or detecting sentence-level inconsistencies does not explicitly require security reasoning about protocol behavior in 3GPP specifications and adversarial exploitation conditions. In contrast, \bench directly evaluates specification-driven security reasoning by incorporating real-world vulnerability scenarios and cross-clause evidence integration.

%% file: latex/benchmark.tex
\section{\bench}

\label{section:bench}
We develop a three-stage benchmark in which each stage introduces increasingly challenging requirements. Stage 1 targets clause-local understanding. Each question is answerable from a single clause and does not require integrating information across clauses, tables, or figures. Stage 2 adds an explicit evidence grounding requirement to enforce verifiability, requiring models to cite the supporting clause/table/figure identifiers. Stage 3 consolidates the remaining challenges by requiring cross-clause integration and security-aware interpretation. Table~\ref{tab:coverage_summary} in Appendix~\ref{appendix:coverage_summary} summarizes the coverage of the external benchmarks incorporated in each stage, including their task formats and the corresponding capabilities they assess. \bench, including the verified and corrected tasks from other datasets, is publicly available~\cite{2026cellularspecsecbench_dataset}. 

\noindent \textbf{Benchmark Scopes.}
\bench is designed to evaluate not only whether models can read 3GPP specifications, but also whether they can perform standards-driven security reasoning. Accordingly, questions span two scopes. \textbf{Specification understanding} tasks target definitions, message formats, parameter constraints, and normative requirements, and test whether a model can correctly interpret standard descriptions. \textbf{Security reasoning} tasks go further by requiring models to reason about how specification-defined procedures and protections can fail under adversarial settings, such as when a message lacks integrity protection, a state transition can be triggered without authentication, or a precondition is underspecified. These tasks evaluate vulnerability-oriented reasoning and require evidence-grounded conclusions linked to the relevant multimodal context.

\subsection{Task Format Definition}
\label{sec:task_def}

We formalize the tasks that involve comprehending 3GPP specifications and identifying security design vulnerabilities into three types. These task types correspond to a progressive evaluation of specification understanding capabilities, ranging from basic comprehension to evidence grounding and finally to security reasoning. Examples for each task type are provided in Appendix~\ref{appendix:task_types}.

\noindent \textbf{Open-ended answering.} It is formalized as 
\(
f_{\text{open}} : (Q) \rightarrow (A, E),
\)
where $Q$ is a query relevant to the 3GPP specification corpus, $A$ is a textual answer, and $E$ is an evidence set consisting of clause numbers, table identifiers, or figure labels. In evidence-grounded settings, $E$ is required to explicitly support the answer.

\noindent \textbf{Multiple-choice answering.} It is defined as 
\(
f_{\text{mc}} : (Q, \mathcal{O}) \rightarrow (A, E),
\)
where $\mathcal{O}$ is a fixed set of candidate options, and exactly one option is compliant with the 3GPP specification. 
In evidence-grounded settings, the model is expected to return both the correct option and $E$.

\noindent \textbf{Vulnerability/inconsistency Labeling.} Formally, it is defined as 
\(
f_{\text{vul}} : S \rightarrow (\mathbf{y}, {E}),
\qquad 
\mathbf{y} = (y_{\text{label}}, y_{\text{multi-label}}),
\)
where $S$ is a normative sentence and $y$ is the label assigned for $S$. There are two main types of labels: 
$y_{\text{label}}$ indicates whether the sentence contains a vulnerability or contradicts other parts of specifications, reflecting a semantic or procedural inconsistency; $y_{\text{multi-label}}$ is a multi-label vulnerability category assignment (e.g., denial-of-service, replay, downgrade, privacy/tracking, spoofing, authentication bypass, etc.). In evidence-grounded settings, $E$ identifies the specification clauses, tables, or figures needed to justify the assigned labels.

\subsection{\bench Construction Steps}
\label{subsect:bench-construction}

Figure~\ref{fig:pipeline} provides an overview of \bench construction steps for two dataset groups:

\begin{figure}[t]
    \centering
    \includegraphics[width=1\linewidth]{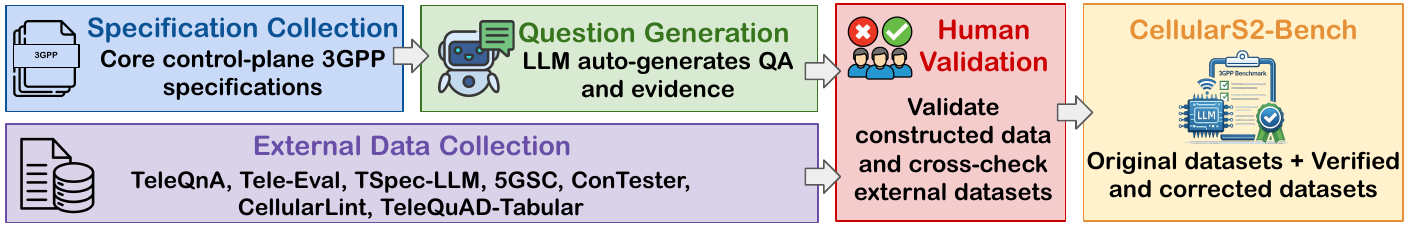}
    \caption{\bench development steps}
    \label{fig:pipeline}
    \vspace{-0.25in}
\end{figure}

\noindent \textbf{Original datasets.} These datasets are developed and constructed from core control-plane 3GPP specifications (see Appendix~\ref{appendix:description_of_specification} for details) that govern signaling procedures, state transitions, and protection mechanisms in 4G and 5G networks. 
They are also where many real-world security issues originate, 
including downgrade, replay, and denial-of-service attacks, often due to insecure transitions, missing integrity protection, or ambiguous normative requirements. To develop datasets, we adopt a semi-automated pipeline combining question-answer (QA) pair generation using large language models (LLMs) (see Figure \ref{fig:Question_Generation} in Appendix \ref{appendix:question_generation}) and expert verification (see Section~\ref{subsect:human-verification}).

\noindent \textbf{Verified and corrected datasets.} To broaden coverage and support diverse task formats, \bench selectively integrates state-of-the-art benchmarks from the telecom and cellular NLP literature. 
It includes TeleQnA, 
Tele-Eval and TSpec-LLM for assessing how models understand 3GPP specifications; it incorporates SPEC5G for 5G text security classification; ConTester for semantic-equivalence sentence pairs for 4G; CellularLint for NLI-style sentence pairs for 4G/5G, and TeleQuAD-Tabular for tabular question answering over specification tables. 
We exclude datasets that are not centered on specification interpretation or standards-driven reasoning (e.g., TeleMath~\cite{colle2025telemathbenchmarklargelanguage}, which primarily targets mathematical problem solving). 
\begin{wrapfigure}{r}{0.42\linewidth}
    \centering
    \vspace{-0.12in}
    \includegraphics[width=\linewidth]{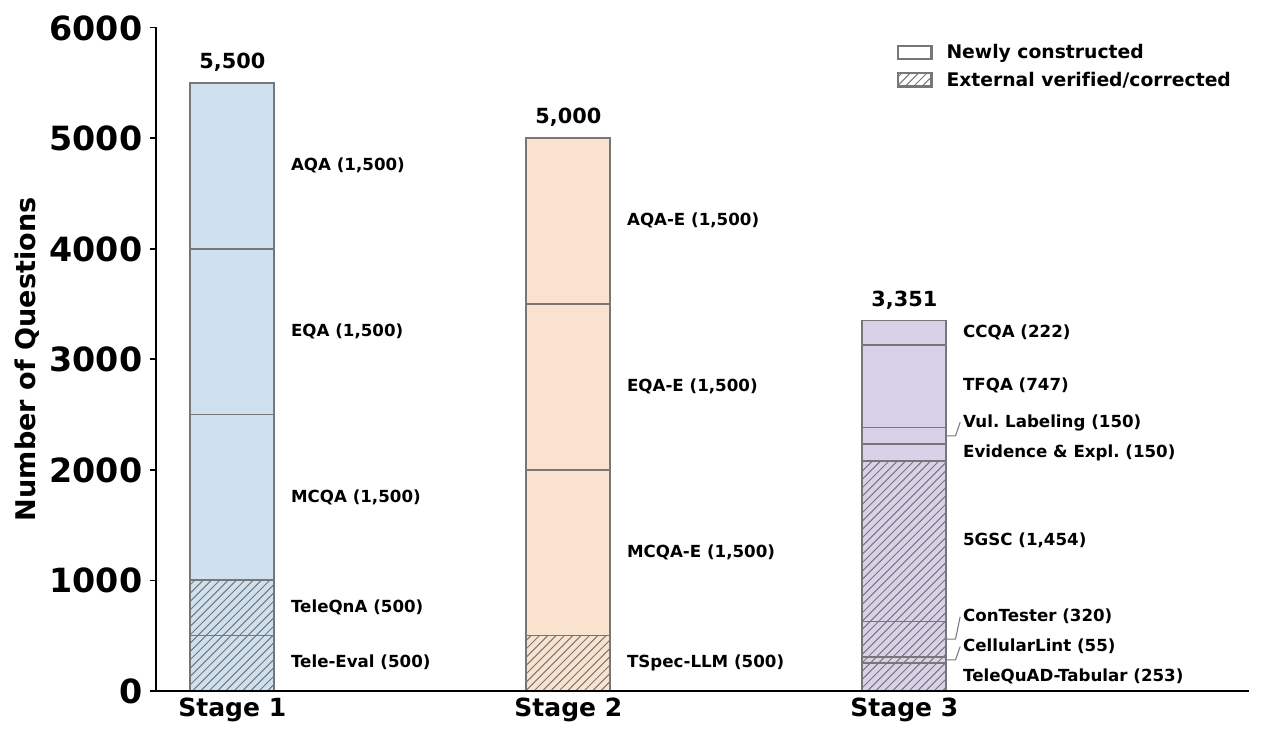}
    \vspace{-0.11in}
    \caption{\bench composition}
    \label{fig:benchmark_stats}
    \vspace{-0.05in}
\end{wrapfigure}
Notably, external datasets are not always expert-validated. 
For example, Tele-Eval, TeleQnA and TSpec-LLM are mainly constructed by LLM-generated entries. We observe that a significant number of entries are inaccurate from an expert's perspective. Consequently, \bench does not ingest these datasets verbatim. Instead, we curate, verify, and correct the integrated subsets, ensuring that each retained instance is consistent with specification evidence and intended task definition. Through this selective integration and expert verification, \bench combines standards-grounded reasoning with broader telecom knowledge, resulting in a benchmark that is more multifaceted and representative than prior datasets in this domain.


\noindent \textbf{\bench Statistics.}
\bench contains 13,851 instances, including 10,269 (74.14\%) newly constructed instances and 3,582 (25.86\%) instances adapted from seven external datasets through expert verification and correction.
Stage~1 contains 5,500 instances; there are 4,500 newly constructed questions, including open-ended answering and multiple-choice questions. Two verified and corrected tasks from TeleQnA and Tele-Eval contribute a total of 1,000 questions. Stage~2 contains 5,000
instances and requires returning the evidence from specifications. 
There are a total of 4,500 newly constructed evidence-grounded questions. Stage~2 also incorporates the verified and corrected external dataset from TSpec-LLM, contributing 500 questions. Stage~3 contains 3,351 instances and focuses on cross-clause integration and security-aware interpretation. Its 1,269 newly constructed instances comprise 222 Cross-Clause Question Answering (CCQA) and 747 Table-and-Figure Question Answering (TFQA). For security reasoning, Stage~3 includes 300 vulnerability/inconsistency labeling instances with complete evidence and explanations. 
The remaining 2,082 instances are externally verified and corrected datasets from 5GSC, ConTester, CellularLint, and TeleQuAD-Tabular, contributing a total of 1,454 sentence-level questions, 320 sentence-pair questions, 55 questions, and 253 table questions. 
Figure~\ref{fig:benchmark_stats} shows the distribution across stages. 
Details of the instances are elaborated in Appendix~\ref{appendix:stage1-3}.

\subsection{Human Verification}
\label{subsect:human-verification}

We ensure dataset reliability through expert-driven annotation and auditing. All datasets are constructed or corrected by domain experts following a consistent verification procedure. Each instance is constructed or verified by one annotator and independently reviewed by two auditors; it is accepted only when both auditors mark it as \textit{Pass}. For external datasets, we verify correctness against specification text, ensure evidence traceability, and exclude instances that are inconsistent or insufficiently grounded. Across all stages, we observe high acceptance rates 97.6\%–99.9\% and strong inter-annotator agreement $\kappa$ ranging from 0.64 to 0.80, indicating high dataset reliability. Detailed annotation procedures and annotator
qualifications are provided in Appendix~\ref{appendix:annotation}.

%% file: latex/methods.tex
\section{\ari}
\label{section:framework}

\begin{figure*}[t]
    \centering
    \includegraphics[width=0.9\textwidth]{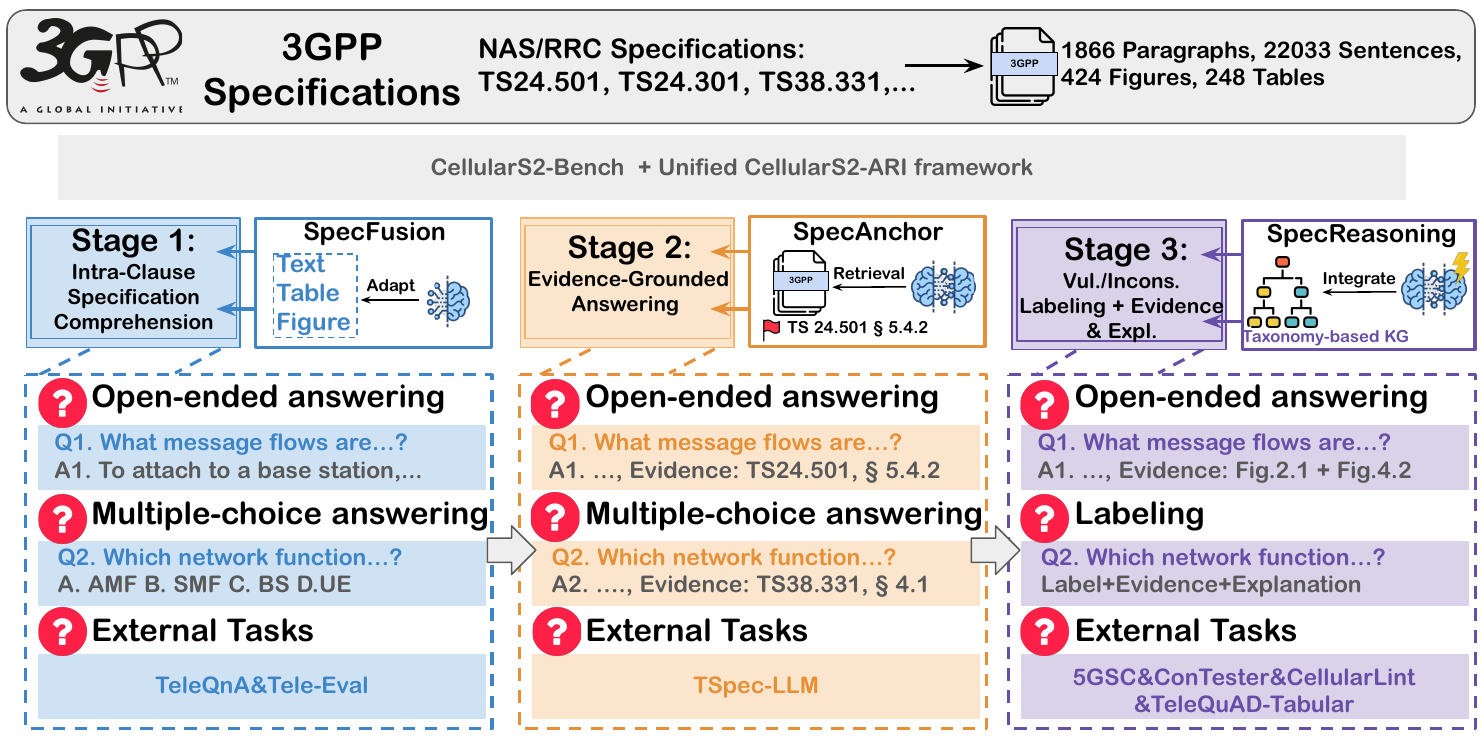}
    \caption{Overview of \ari and \bench}
    \label{fig:Overview}
    \vspace{-0.25in}
\end{figure*}

In this section, we present a unified framework \ari. We apply \ari to \bench as a baseline framework to enable reproducible and systematic security analysis of 3GPP specifications.  
This three-module framework is inspired by the Dreyfus model of skill acquisition~\cite{10.5555/7916}: the \textit{Adapt} corresponds to novice-level rule-based comprehension, the \textit{Retrieve} aligns with competent, evidence-guided decision making, and the \textit{Integrate} reflects expert-level synthesis across multiple sources.  The implementation of the framework is publicly available on our website \cite{CellSpecSecBench}. 

\noindent \textbf{Flexible Framework.} Importantly, \ari is designed as a capability-oriented framework rather than a fixed system. It specifies what each module should accomplish, but does not prescribe how that capability must be implemented. Thus, each layer is implementation-agnostic and can be instantiated with different parsers, retrievers, or knowledge representations for advanced performance.

As shown in Figure~\ref{fig:Overview}, \ari is composed of three modules: \textbf{\textit{SpecFusion}} (\underline{Spec}ification \underline{Fusion}), \textbf{\textit{SpecAnchor}} (\underline{Spec}ification \underline{Anchor}), and \textbf{\textit{SpecReasoning}} (\underline{Spec}ifications Knowledge Integration \underline{Reasoning}), which progressively function as the abilities to \textit{adapt} 3GPP specifications, \textit{retrieve} knowledge from different standards, and \textit{integrate} this knowledge to conduct security analysis on 3GPP specifications. \ari requires no task-specific fine-tuning and doesn't access
benchmark answers, labels, or gold evidence during retrieval or answer generation. \textbf{The design and implementation described below instantiate one prototype instead of imposing a set of static mechanisms.}

\subsection{Adaptation - \textit{SpecFusion}}
\textit{SpecFusion} is designed to enable \ari to \textit{adapt} to the domain knowledge in 3GPP specifications. LLMs trained primarily on open-domain corpora often struggle to interpret the stylistic and structural domain-specific conventions of technical standards. In particular, they tend to (1) misinterpret domain-specific text for hallucinations and (2) mishandle multimodal contents, including tables and figures for depicting concepts that are hard to describe in natural language (e.g., transitions between multiple steps).

Formally, each 3GPP specification can be decomposed into 3 components \(
\mathcal{C} = 
\mathcal{C}_{\text{text}} 
\cup 
\mathcal{C}_{\text{table}} 
\cup 
\mathcal{C}_{\text{figure}},
\)
Textual components ($\mathcal{C}_{\text{text}}$) correspond to normative paragraphs written in natural language that describe procedures, operational conditions, definitions, or constraints. In 3GPP specifications, these paragraphs define how entities behave in different conditions and settings. Such text chunks provide \ari with fine-grained exposure to the wording and normative style of standardization text. Table components ($\mathcal{C}_{\text{table}}$) represent normative tables that encode structured relationships among parameters, states, timers, configuration options, and actions. Including tabular chunks in the adaptation process allows \ari to learn how rules and mappings are expressed in the tables of 3GPP specifications. Figure components ($\mathcal{C}_{\text{figure}}$) correspond to diagrams such as signaling flows, state machines, and message or information-element formats. 
Including figure components in the adaptation process provides \ari with the ability to interpret protocol information that is expressed graphically in specifications.

\noindent \textbf{Multimodal adaptation.} Rather than applying a separate encoder to each modality, our prototype transforms non-textual components into a unified textual representation. Specifically, tables and figures in 3GPP specifications are extracted and converted into structured JSON descriptions, as described in Appendix~\ref{appendix:content_extraction}, resulting in a unified textual interface for heterogeneous specification content. It allows textual clauses, tables, and figures to be represented within the same component space $\mathcal{C}$. 
Unlike many prior works that remove or ignore multimodal components from 3GPP specifications~\cite{karim2023spec5gdataset5gcellular, 298168}, \textit{SpecFusion} preserves tables and figures, enabling the adaptation to multimodal knowledge in standards. 

The role of \textit{SpecFusion} is therefore to normalize and preserve specification knowledge for the simple intra-clause specification comprehension, not to determine which evidence is most relevant to a particular query. Relevance ranking and evidence selection are handled by the subsequent \textit{SpecAnchor} for evidence-grounded answering.

\subsection{Retrieval - \textit{SpecAnchor}}
Since each 3GPP specification can span hundreds to thousands of pages and there are thousands of specifications for operational cellular networks, relying solely on a model's parametric memory is insufficient for trustworthy standards analysis. 
\textit{SpecAnchor} addresses this challenge by performing retrieval over the unified component space constructed by \textit{SpecFusion}. Its goal is to anchor model outputs to explicit evidence from 3GPP specifications, thereby improving verifiability and supporting downstream security reasoning. Formally, for each query $Q$, the \textit{SpecAnchor} searches the retrieval database constructed from the multimodal components produced by \textit{SpecFusion}. It returns a set of relevant evidence chunks denoted as $E(Q) = \{(c_j, \ell_j)\}_{j=1}^{m}$, where each chunk $c_j$ corresponds to a subclause, table, or figure, and $\ell_j$ is its citation label (e.g., clause number). This set is further provided to the framework to ensure each answer, decision, or security conclusion can be explicitly grounded in 3GPP specifications.

In our prototype, \textit{SpecAnchor} uses a hybrid retrieval strategy. The sparse component $s_{\text{sp}}(Q,c)$ is computed using a BM25 ranker that captures exact keyword matches between the query $q$ and candidate chunk $c$. The dense component $s_{\text{de}}(Q,c)$ is obtained by encoding the query and each chunk into TF--IDF vectors, projecting them into a lower-dimensional latent space via truncated SVD, and computing cosine similarity between the resulting vectors. These dense embeddings approximate semantic similarity using statistical co-occurrence patterns and are efficiently indexed for similarity search. The final relevance score is a weighted combination:
\(s(Q,c) = \alpha\, s_{\text{sp}}(Q,c) + (1-\alpha)\, s_{\text{de}}(Q,c), \hfill \alpha \in [0,1],\)
where $\alpha$ is tuned on a small validation set to balance lexical precision and semantic recall. The top-$k$ highest-scoring chunks form the Retrieved Evidence context supplied to the model. See Appendix~\ref{appendix:top-k} for the selection analysis of $k$.

\noindent \textbf{Sample query and output.} When queried with ``During Emergency Registration, when is NAS integrity established?'', \textit{SpecAnchor} retrieves multimodal evidence including TS 24.501, Clause 5.5.1 Registration procedure and Clause 5.4.2 Security Mode Control procedure. The former describes how the UE initiates registration, while the latter specifies when NAS integrity protection is activated. Concatenating these clauses yields the evidence-grounded context, from which \ari generates the trustworthy answer: ``NAS integrity is activated through Security Mode Control; emergency registration may instead use the null integrity algorithm 5G-IA0.'' accompanied by explicit citations: TS 24.501, Clauses 5.5.1 and 5.4.2.

\subsection{Integration - \textit{SpecReasoning}}

While prior two modules expose 3GPP specifications to \ari and enable it to retrieve multimodal evidence to generate evidence-grounded trustworthy outputs, the pieces of knowledge from the clause, table, and figure have not been integrated together for performing complex reasoning. Likewise, cross-modal relationships among textual, tabular, and visual evidence are often obscured, since tabular semantics depend on multi-row conditions and figures encode procedural or causal flows that cannot be captured through text retrieval alone. To overcome these limitations and enable relational reasoning, we transform the multimodal outputs of \textit{SpecAnchor} into structured, interlinked representations. Specifically, we convert them into a taxonomy-based knowledge graph (KG) that explicitly models the structural and procedural dependencies among 3GPP specification entities.

\noindent \textbf{Entity types for taxonomy-based KG.} We first construct an entity taxonomy that reflects the hierarchical organization of 3GPP specifications. By systematically analyzing each clause, including all multimodal components, we identify the recurring entities from the majority of clauses and summarize a set of entity types as shown in Appendix~\ref{append:Taxonomy-based KG} Table~\ref{tab:entity-types}: signaling procedures, messages, information elements, identifiers, timers, states, conditions, and system-level properties. They have consistent meanings across 3GPP specifications and form the basic entity vocabulary used in taxonomy-based KGs.

\noindent \textbf{Relation types for taxonomy-based KG.} We next derive the relation schema that shows how these entities interact in 3GPP specifications. Two complementary families of relations are defined. \textbf{Core relations} $\mathcal{R}_{\text{core}}$ represent the structural dependencies that govern protocol execution logic. These relations describe, for example, how a procedure is decomposed into signaling steps, how a message contains a set of information elements, how the execution of a step requires the device to be in a specific protocol state, and how timers are started, stopped, or reset during the procedure. They capture the functional organization of the specification independently of whether the behavior has security relevance.  \textbf{Security relations} $\mathcal{R}_{\text{sec}}$ capture dependencies that explicitly relate to protection requirements and security outcomes. These relations encode whether a message or information element must be integrity protected or ciphered, whether an exception exists in which an otherwise protected object may appear in unprotected form, whether the processing of a message is permitted only after an integrity check is validated, and whether the reception of a message triggers a network action or induces a transition in the protocol state machine. Appendix~\ref{append:Taxonomy-based KG} Table~\ref{tab:relation-types} summarizes the full inventory of relation types used in \textit{SpecReasoning}.

\noindent \textbf{Building taxonomy-based KG.} Given the entity and relation types, \textit{SpecReasoning} instantiates the KGs with the retrieved evidence. Formally, a labeled multi-relational graph is denoted as \(G = (V, E, \tau, \rho, \lambda)\), where $V$ is the set of nodes and $E \subseteq V \times V$ is the set of directed edges. Each node in $V$ is assigned an entity type $\tau(v)$ chosen from the entity table {Procedure, Message, \ldots{}, Property}. Each edge in $E$ is assigned a relation type $\rho(e)$ drawn from $\mathcal{R}{\text{core}} \cup \mathcal{R}{\text{sec}}$. $\lambda$ stores the provenance metadata for each node and edge, including its defining clause, table, or figure identifier (e.g., ``TS 24.501, Clause X.Y.Z''). See Appendix~\ref{append:Taxonomy-based KG} for the example of a taxonomy-based KG.

\noindent \textbf{Taxonomy-based KG enhanced reasoning.} We use the taxonomy-based KGs as the additional retrievable evidence source. Each clause-level KG is converted into a text block, which includes its clause identifier, the extracted entities, relations, and the original text in specifications. The same hybrid retrieval strategy used in \textit{SpecAnchor} is applied over the KG-derived text blocks to retrieve top-$k$ relevant KG blocks to help with reasoning correctness.

%% file: latex/evaluation.tex
\section{Evaluation}
\label{section:evaluation}

\noindent \textbf{Implementation and Configuration.}
We deploy the prototype \ari on a remote Jetstream2 server~\cite{jetstream2025llminference} equipped with an H100 GPU. To evaluate the generality of \ari, we conduct experiments across six representative models: GPT-5.6 Terra \cite{openai_gpt56_2026}, Claude Sonnet 5 \cite{anthropic_sonnet5_2026}, DeepSeek-V3.2-Exp~\cite{deepseek_v3_2_exp_2025}, Qwen-plus-2025-12-01~\cite{qwen_plus_2025}, Llama-4-Scout-17B-16E-Instruct~\cite{llama4_scout_17b_2025}, and Qwen3-30B-A3B-Instruct-2507~\cite{qwen3_30b_a3b_2025}. To ensure deterministic and reproducible results, we set the model temperature to 0 for all experiments. 
For each model, we evaluate two configurations: (1) \textbf{Base}, which uses the original model without any \ari modules, and (2) \textbf{\ari}, which augments the same underlying model with the full \ari pipeline, including \textit{SpecFusion}, \textit{SpecAnchor}, and \textit{SpecReasoning}. It allows us to measure the performance gains introduced by \ari independently of the choice of backbone model. Implementation details are provided in Appendix~\ref{appendix:implementation_details}.

\begin{table}[p]  
\caption{Overall Task Performance across Six Models}
\label{tab:overall_task_results}
\centering

\fontsize{6}{6.5}\selectfont   
\setlength{\tabcolsep}{2pt}
\renewcommand{\arraystretch}{0.9}

\begin{adjustbox}{max width=\linewidth,max height=0.88\textheight}
\begin{tabular}{lllcccc}
\hline
\textbf{Model} & \textbf{Task} & \textbf{Metric} & \textbf{BASE} & \textbf{\textit{SpecFusion}} & \textbf{+\textit{SpecAnchor}} & \textbf{\ari} \\
\hline
\rowcolor{stage1color}
\multicolumn{7}{c}{\textbf{Stage 1 -- Intra-Clause
Specification Comprehension}}\\
\hline
GPT-5.6 Terra & EQA & Score=2 & 35.73\% & 81.13\% & 86.60\% & 86.60\% \\
GPT-5.6 Terra & AQA & Score=2 & 32.00\% & 75.47\% & 70.60\% & 80.00\% \\
GPT-5.6 Terra & MCQA & Accuracy & 90.67\% & 90.27\% & 90.07\% & 100.00\% \\

Claude Sonnet 5 & EQA & Score=2 & 34.60\% & 83.80\% & 88.20\% & 88.20\% \\
Claude Sonnet 5 & AQA & Score=2 & 32.00\% & 77.13\% & 83.20\% & 83.20\% \\
Claude Sonnet 5 & MCQA & Accuracy & 88.67\% & 88.40\% & 88.20\% & 88.60\% \\

DeepSeek-V3.2-Exp & EQA & Score=2 & 35.87\% & 85.80\% & 86.25\% & 97.73\% \\
DeepSeek-V3.2-Exp & AQA & Score=2 & 34.00\% & 81.40\% & 81.82\% & 97.00\% \\
DeepSeek-V3.2-Exp & MCQA & Accuracy & 76.00\% & 73.40\% & 93.38\% & 100.00\% \\

Qwen3-Plus & EQA & Score=2 & 31.60\% & 82.93\% & 87.00\% & 88.40\% \\
Qwen3-Plus & AQA & Score=2 & 30.53\% & 87.60\% & 88.96\% & 89.40\% \\
Qwen3-Plus & MCQA & Accuracy & 79.40\% & 84.00\% & 93.38\% & 96.00\% \\

Llama4-Scout-17B & EQA & Score=2 & 4.33\% & 65.60\% & 66.00\% & 66.93\% \\
Llama4-Scout-17B & AQA & Score=2 & 2.07\% & 57.33\% & 57.60\% & 59.80\% \\
Llama4-Scout-17B & MCQA & Accuracy & 78.27\% & 78.33\% & 78.00\% & 78.53\% \\

Qwen3-30B & EQA & Score=2 & 31.87\% & 73.00\% & 76.33\% & 76.67\% \\
Qwen3-30B & AQA & Score=2 & 22.00\% & 63.73\% & 63.73\% & 66.60\% \\
Qwen3-30B & MCQA & Accuracy & 69.87\% & 69.87\% & 69.20\% & 70.13\% \\
\hline
\rowcolor{stage1color}
\multicolumn{7}{c}{\textbf{Stage 1 -- Verified and Corrected Tasks}}\\
\hline
GPT-5.6 Terra & TeleQnA & Accuracy & 95.20\% & 94.00\% & 96.00\% & 96.00\% \\
GPT-5.6 Terra & Tele-Eval & Score=2 & 41.20\% & 39.60\% & 98.00\% & 98.00\% \\

Claude Sonnet 5 & TeleQnA & Accuracy & 85.20\% & 86.40\% & 90.40\% & 90.40\% \\
Claude Sonnet 5 & Tele-Eval & Score=2 & 46.20\% & 68.00\% & 97.00\% & 98.00\% \\

DeepSeek-V3.2-Exp & TeleQnA & Accuracy & 90.00\% & 94.20\% & 94.00\% & 96.00\% \\
DeepSeek-V3.2-Exp & Tele-Eval & Score=2 & 38.00\% & 97.40\% & 98.00\% & 97.40\% \\

Qwen3-Plus & TeleQnA & Accuracy & 23.00\% & 85.80\% & 86.00\% & 96.80\% \\
Qwen3-Plus & Tele-Eval & Score=2 & 14.00\% & 62.80\% & 92.00\% & 93.20\% \\

Llama4-Scout-17B & TeleQnA & Accuracy & 85.20\% & 84.40\% & 86.00\% & 90.20\% \\
Llama4-Scout-17B & Tele-Eval & Score=2 & 17.20\% & 65.80\% & 60.60\% & 76.60\% \\

Qwen3-30B & TeleQnA & Accuracy & 88.00\% & 92.00\% & 92.00\% & 94.00\% \\
Qwen3-30B & Tele-Eval & Score=2 & 25.20\% & 71.80\% & 71.40\% & 80.00\% \\

\hline
\rowcolor{stage2color}
\multicolumn{7}{c}{\textbf{Stage 2 -- Evidence-Grounded Answering}}\\
\hline
GPT-5.6 Terra & EQA-E & Score=2 / Evidence & 36.60\%/0.00\% & 87.27\%/100.00\% & 87.27\%/100.00\% & 89.47\%/99.73\% \\
GPT-5.6 Terra & AQA-E & Score=2 / Evidence & 31.80\%/0.00\% & 79.93\%/100.00\% & 79.93\%/100.00\% & 83.87\%/100.00\% \\
GPT-5.6 Terra & MCQA-E & Accuracy / Evidence & 89.73\%/0.00\% & 98.33\%/100.00\% & 98.53\%/100.00\% & 100.00\%/100.00\% \\

Claude Sonnet 5 & EQA-E & Score=2 / Evidence & 35.73\%/0.00\% & 89.33\%/99.73\% & 86.60\%/100.00\% & 86.60\%/100.00\% \\
Claude Sonnet 5 & AQA-E & Score=2 / Evidence & 30.53\%/0.00\% & 85.33\%/100.00\% & 83.00\%/100.00\% & 83.00\%/100.00\% \\
Claude Sonnet 5 & MCQA-E & Accuracy / Evidence & 88.87\%/0.00\% & 97.53\%/100.00\% & 98.33\%/100.00\% & 100.00\%/100.00\% \\

DeepSeek-V3.2-Exp & EQA-E & Score=2 / Evidence & 30.00\%/0.00\% & 90.00\%/0.00\% & 96.00\%/97.00\% & 96.80\%/96.40\% \\
DeepSeek-V3.2-Exp & AQA-E & Score=2 / Evidence & 39.20\%/0.00\% & 89.61\%/0.00\% & 96.40\%/96.53\% & 96.80\%/96.40\% \\
DeepSeek-V3.2-Exp & MCQA-E & Accuracy / Evidence & 79.40\%/0.00\% & 94.12\%/0.00\% & 100.00\%/100.00\% & 100.00\%/100.00\% \\

Qwen3-Plus & EQA-E & Score=2 / Evidence & 27.93\%/0.00\% & 89.38\%/0.00\% & 91.40\%/96.47\% & 95.00\%/97.00\% \\
Qwen3-Plus & AQA-E & Score=2 / Evidence & 25.60\%/0.00\% & 79.87\%/0.00\% & 86.47\%/94.00\% & 93.00\%/95.40\% \\
Qwen3-Plus & MCQA-E & Accuracy / Evidence & 80.40\%/0.00\% & 96.32\%/0.00\% & 98.80\%/93.73\% & 98.80\%/95.00\% \\

Llama4-Scout-17B & EQA-E & Score=2 / Evidence & 26.07\%/0.00\% & 65.00\%/0.60\% & 75.27\%/93.73\% & 84.33\%/96.47\% \\
Llama4-Scout-17B & AQA-E & Score=2 / Evidence & 10.47\%/0.00\% & 70.00\%/1.40\% & 72.27\%/93.47\% & 81.47\%/94.00\% \\
Llama4-Scout-17B & MCQA-E & Accuracy / Evidence & 76.87\%/0.00\% & 80.00\%/0.00\% & 86.87\%/93.73\% & 89.80\%/88.40\% \\

Qwen3-30B & EQA-E & Score=2 / Evidence & 23.13\%/0.00\% & 73.60\%/1.47\% & 76.47\%/93.73\% & 83.87\%/96.47\% \\
Qwen3-30B & AQA-E & Score=2 / Evidence & 19.33\%/0.00\% & 66.47\%/2.67\% & 73.53\%/93.47\% & 82.47\%/94.00\% \\
Qwen3-30B & MCQA-E & Accuracy / Evidence & 67.60\%/0.00\% & 92.60\%/0.00\% & 92.07\%/88.40\% & 94.20\%/93.73\% \\
\hline
\rowcolor{stage2color}
\multicolumn{7}{c}{\textbf{Stage 2 -- Verified and Corrected Tasks}}\\
\hline
GPT-5.6 Terra & TSpec-LLM & Accuracy / Evidence & 83.40\%/0.00\% & 95.80\%/72.20\% & 95.80\%/72.20\% & 96.60\%/85.20\% \\
Claude Sonnet 5 & TSpec-LLM & Accuracy / Evidence & 80.40\%/0.00\% & 96.60\%/72.20\% & 95.40\%/85.20\% & 95.40\%/85.20\% \\
DeepSeek-V3.2-Exp & TSpec-LLM & Accuracy / Evidence & 76.00\%/0.00\% & 78.00\%/0.00\% & 94.00\%/88.00\% & 94.00\%/92.00\% \\
Qwen3-Plus & TSpec-LLM & Accuracy / Evidence & 77.40\%/0.00\% & 80.00\%/36.00\% & 95.20\%/88.20\% & 95.20\%/88.20\% \\
Llama4-Scout-17B & TSpec-LLM & Accuracy / Evidence & 69.60\%/0.00\% & 73.60\%/0.00\% & 89.20\%/72.20\% & 92.60\%/88.20\% \\
Qwen3-30B & TSpec-LLM & Accuracy / Evidence & 71.60\%/0.00\% & 92.40\%/0.00\% & 90.60\%/72.20\% & 92.60\%/88.20\% \\
\hline
\rowcolor{stage3color}
\multicolumn{7}{c}{\textbf{Stage 3 -- Vulnerability/Inconsistency Labeling with Evidence and Explanations}}\\
\hline
GPT-5.6 Terra & CCQA & Score=2 / Evidence & 21.17\%/0.00\% & 66.22\%/82.88\% & 66.22\%/82.88\% & 71.17\%/100.00\% \\
GPT-5.6 Terra & TFQA & Score=2 / Evidence & 48.06\%/0.00\% & 55.42\%/0.00\% & 55.42\%/0.00\% & 64.26\%/100.00\% \\
GPT-5.6 Terra & Vulnerability Labeling & Bin/Micro/Macro & 75.41\%/31.30\%/11.13\% & 78.12\%/50.00\%/53.01\% & 80.00\%/70.00\%/73.00\% & 90.00\%/82.88\%/88.00\% \\
GPT-5.6 Terra & Evidence \& Explanations & Correctness & 0.00\% & 0.00\% & 0.00\% & 100.00\% \\

Claude Sonnet 5 & CCQA & Score=2 / Evidence & 19.37\%/0.00\% & 62.61\%/80.18\% & 67.12\%/100.00\% & 90.00\%/100.00\% \\
Claude Sonnet 5 & TFQA & Score=2 / Evidence & 51.54\%/0.00\% & 57.83\%/0.00\% & 61.98\%/100.00\% & 89.02\%/100.00\% \\
Claude Sonnet 5 & Vulnerability Labeling & Bin/Micro/Macro & 46.55\%/16.78\%/8.02\% & 53.19\%/18.03\%/5.79\% & 84.00\%/72.00\%/73.00\% & 88.52\%/81.13\%/86.99\% \\
Claude Sonnet 5 & Evidence \& Explanations & Correctness & 0.00\% & 0.00\% & 100.00\% & 100.00\% \\

DeepSeek-V3.2-Exp & CCQA & Score=2 / Evidence & 17.28\%/0.00\% & 59.07\%/0.00\% & 61.86\%/72.09\% & 95.35\%/91.16\% \\
DeepSeek-V3.2-Exp & TFQA & Score=2 / Evidence & 17.27\%/0.00\% & 43.11\%/41.10\% & 54.75\%/0.00\% & 97.99\%/95.45\% \\
DeepSeek-V3.2-Exp & Vulnerability Labeling & Bin/Micro/Macro & 80.00\%/39.02\%/38.81\% & 80.00\%/33.03\%/25.00\% & 80.00\%/40.00\%/39.00\% & 93.02\%/70.59\%/89.35\% \\
DeepSeek-V3.2-Exp & Evidence \& Explanations & Correctness & 10.00\% & 10.00\% & 53.33\% & 88.00\% \\

Qwen3-Plus & CCQA & Score=2 / Evidence & 12.61\%/0.00\% & 36.36\%/0.00\% & 50.00\%/77.27\% & 90.99\%/90.99\% \\
Qwen3-Plus & TFQA & Score=2 / Evidence & 27.18\%/0.00\% & 44.59\%/0.00\% & 50.00\%/41.89\% & 88.22\%/83.00\% \\
Qwen3-Plus & Vulnerability Labeling & Bin/Micro/Macro & 37.76\%/12.72\%/7.58\% & 69.57\%/25.64\%/18.64\% & 72.73\%/24.00\%/17.15\% & 90.91\%/70.00\%/85.38\% \\
Qwen3-Plus & Evidence \& Explanations & Correctness & 0.00\% & 0.00\% & 53.33\% & 95.33\% \\

Llama4-Scout-17B & CCQA & Score=2 / Evidence & 11.26\%/0.00\% & 48.00\%/0.90\% & 54.00\%/54.90\% & 63.06\%/89.19\% \\
Llama4-Scout-17B & TFQA & Score=2 / Evidence & 27.04\%/0.00\% & 28.51\%/8.03\% & 29.99\%/3.75\% & 63.19\%/77.78\% \\
Llama4-Scout-17B & Vulnerability Labeling & Bin/Micro/Macro & 56.25\%/21.51\%/17.90\% & 50.00\%/23.00\%/17.02\% & 60.00\%/24.00\%/15.98\% & 80.00\%/71.43\%/83.38\% \\
Llama4-Scout-17B & Evidence \& Explanations & Correctness & 0.00\% & 4.00\% & 30.00\% & 62.00\% \\

Qwen3-30B & CCQA & Score=2 / Evidence & 12.61\%/0.00\% & 51.80\%/2.70\% & 58.56\%/54.95\% & 66.67\%/89.19\% \\
Qwen3-30B & TFQA & Score=2 / Evidence & 24.36\%/0.00\% & 33.60\%/3.88\% & 30.39\%/11.78\% & 64.52\%/77.78\% \\
Qwen3-30B & Vulnerability Labeling & Bin/Micro/Macro & 28.57\%/34.39\%/44.98\% & 62.07\%/35.41\%/47.01\% & 76.00\%/36.28\%/48.19\% & 78.00\%/71.40\%/83.40\% \\
Qwen3-30B & Evidence \& Explanations & Correctness & 0.00\% & 6.67\% & 30.00\% & 78.00\% \\

\hline
\rowcolor{stage3color}
\multicolumn{7}{c}{\textbf{Stage 3 -- Verified and Corrected Tasks}}\\
\hline
GPT-5.6 Terra & 5GSC & Accuracy & 93.40\% & 94.02\% & 100.00\% & 100.00\% \\
GPT-5.6 Terra & ConTester & Accuracy & 78.14\% & 80.05\% & 76.50\% & 79.78\% \\
GPT-5.6 Terra & CellularLint & Accuracy & 40.00\% & 54.55\% & 98.18\% & 100.00\% \\
GPT-5.6 Terra & TeleQuAD-Tabular & Score=2 / Evidence & 18.69\%/0.00\% & 67.29\%/0.00\% & 83.64\%/100.00\% & 100.00\%/100.00\% \\

Claude Sonnet 5 & 5GSC & Accuracy & 40.03\% & 75.86\% & 98.01\% & 100.00\% \\
Claude Sonnet 5 & ConTester & Accuracy & 73.50\% & 78.96\% & 98.09\% & 100.00\% \\
Claude Sonnet 5 & CellularLint & Accuracy & 45.45\% & 54.55\% & 100.00\% & 100.00\% \\
Claude Sonnet 5 & TeleQuAD-Tabular & Score=2 / Evidence & 21.03\%/0.00\% & 67.29\%/0.00\% & 85.05\%/100.00\% & 100.00\%/100.00\% \\

DeepSeek-V3.2-Exp & 5GSC & Accuracy & 46.00\% & 47.00\% & 47.00\% & 100.00\% \\
DeepSeek-V3.2-Exp & ConTester & Accuracy & 96.00\% & 88.89\% & 91.67\% & 100.00\% \\
DeepSeek-V3.2-Exp & CellularLint & Accuracy & 36.00\% & 40.00\% & 40.00\% & 100.00\% \\
DeepSeek-V3.2-Exp & TeleQuAD-Tabular & Score=2 / Evidence & 15.42\%/0.00\% & 64.49\%/0.00\% & 78.06\%/91.37\% & 78.06\%/91.37\% \\

Qwen3-Plus & 5GSC & Accuracy & 66.78\% & 64.29\% & 64.29\% & 96.01\% \\
Qwen3-Plus & ConTester & Accuracy & 69.67\% & 80.56\% & 83.33\% & 100.00\% \\
Qwen3-Plus & CellularLint & Accuracy & 40.00\% & 40.00\% & 40.00\% & 100.00\% \\
Qwen3-Plus & TeleQuAD-Tabular & Score=2 / Evidence & 14.02\%/0.00\% & 64.95\%/0.00\% & 93.46\%/100.00\% & 93.46\%/100.00\% \\

Llama4-Scout-17B & 5GSC & Accuracy & 63.55\% & 61.00\% & 61.21\% & 82.32\% \\
Llama4-Scout-17B & ConTester & Accuracy & 78.69\% & 78.69\% & 78.69\% & 81.97\% \\
Llama4-Scout-17B & CellularLint & Accuracy & 38.18\% & 40.00\% & 40.00\% & 72.73\% \\
Llama4-Scout-17B & TeleQuAD-Tabular & Score=2 / Evidence & 13.08\%/2.80\% & 63.55\%/0.00\% & 76.17\%/94.86\% & 76.17\%/94.86\% \\

Qwen3-30B & 5GSC & Accuracy & 76.89\% & 77.03\% & 77.03\% & 87.00\% \\
Qwen3-30B & ConTester & Accuracy & 69.95\% & 69.13\% & 69.67\% & 83.33\% \\
Qwen3-30B & CellularLint & Accuracy & 38.18\% & 40.00\% & 43.64\% & 81.82\% \\
Qwen3-30B & TeleQuAD-Tabular & Score=2 / Evidence & 11.20\%/0.00\% & 59.81\%/0.00\% & 80.37\%/94.86\% & 80.37\%/94.86\% \\

\bottomrule
\end{tabular}
\end{adjustbox}
\end{table}

\noindent \textbf{Evaluation Metrics.}  For each type of tasks, we evaluate the performance using the following metrics:

$\diamond$ \textbf{Open-ended answering.} We evaluate the answer correctness using an LLM-as-judge protocol with a three-level rubric: 2 (fully correct), 1 (partially correct), and 0 (incorrect). The judge is provided with (i) the question, (ii) the model's answer, and (iii) the gold reference answer. 
To reduce reliance on a single evaluator, we use vanilla DeepSeek-V3.2-Exp as the judge for DeepSeek-V3.2-Exp,
Qwen-plus-2025-12-01, Llama-4-Scout-17B-16E-Instruct, and
Qwen3-30B-A3B-Instruct-2507, and vanilla Claude Sonnet
4.6~\cite{anthropic_sonnet46_2026} as the judge for GPT-5.6 Terra and Claude Sonnet 5. 
We run the judge with temperature 0 and a single decoding to ensure deterministic scoring. To reduce bias, we hide system identities and randomize the presentation order of candidate answers. 
For tasks that require evidence, it is provided with
(iv) the gold evidence, including clause numbers, table/figure identifiers,
and the corresponding extracted content. Evidence correctness is evaluated separately by matching the evidence
identifiers associated with the retrieved results against the gold dataset.
The method to evaluate evidence correctness is described in Appendix~\ref{appendix:evidence_correctness}. The full prompt template and the analysis of LLM-as-judge reliability are illustrated in Appendix~\ref{appendix:human_validation}. 

$\diamond$ \textbf{Multiple-choice answering.} For this task, we report accuracy, since each question has exactly one specification-compliant option. When evidence is required, we also report evidence correctness.

$\diamond$ \textbf{Labeling tasks.} For vulnerability/inconsistency labeling, we evaluate outputs hierarchically. For the binary label (vulnerable/inconsistent vs.\ non-vulnerable), we report positive-class F1. For the multi-label vulnerability categories, we evaluate only on instances with gold positives and report Micro-F1 and Macro-F1. For the evidence and explanations, we treat evidence correctness as the primary metric, since the task requires retrieving the complete set of clauses/tables/figures that collectively justify why a vulnerability arises. We score the explanation using the aforementioned LLM-as-judge approach.

\noindent$\diamond$ \textbf{External Tasks.} For verified and corrected external tasks, we apply the same task-specific metrics after curation: accuracy for multiple-choice, the LLM-as-judge rubric for open-ended answering, and the above classification metrics for labeling tasks.

\noindent \textbf{Overall Performance.} Table~\ref{tab:overall_task_results} 
report the performance of \ari on \bench. We highlight four key observations. 
\textbf{(1) Performance gains from \ari modules.} The base model performs poorly on most tasks, indicating that the base model struggles with the normative style, precision, and logic in 3GPP specifications. Incorporating \ari modules substantially improves performance, demonstrating that domain adaptation and structured reasoning improve standards comprehension and reasoning capability. 
With \ari enabled, we observe better performance on all tasks.
\textbf{(2) Effective evidence grounding via retrieval augmentation.} The base model has poor performance when tasks require evidence. Enabling \textit{SpecAnchor} improves both answer quality and evidence correctness by retrieving and anchoring responses to relevant clauses, tables, and figures, leading to large gains on evidence-grounded tasks.  
\textbf{(3) Strong integration reasoning for security analysis.} Stage~3 represents the most challenging setting, requiring cross-clause integration, vulnerability/inconsistency labeling with complete evidence and explanations. In this difficult setting, strong performance is observed when \ari is applied.
\textbf{(4) Robust improvements on verified and corrected external tasks.} On all verified and corrected external tasks, \ari consistently outperforms the base model, showing that its benefits generalize beyond our newly constructed datasets.

\textbf{Error Analysis.} We analyze the incorrect answers and discuss the potential improvements.

$\diamond$ \textbf{Answer--evidence mismatch.} \ari sometimes generates a correct answer but cites an incorrect clause/table/figure as the evidence. This typically occurs because 3GPP specifications contain multiple semantically related components that share near-identical terminology (e.g., repeated definitions across procedures), making it easy for retrieval a plausible but incorrect reference. This mismatch can be mitigated by strengthening evidence selection and verification, such as using clause-aware reranking that jointly optimizes answer relevance and citation, or adding a citation consistency check that validates whether the cited unit explicitly contains the key conditions/terms.

$\diamond$ \textbf{Evidence incompleteness in security reasoning and labeling.} For security-focused tasks, vulnerability/inconsistency labeling, the dominant failures stem from incomplete evidence, rather than purely incorrect security reasoning. Explanations are directionally correct but incomplete, and the cited evidence set is missing one or more critical dependencies. These patterns emphasize that security reasoning in standards is compositional. To mitigate this, we can add a completeness verifier that rejects conclusions or causal explanations unless the assembled evidence jointly supports the claim.

%% file: latex/limitation.tex
\section{Limitation and Future Work}
\label{section:limitation_and_future_work}
We acknowledge several limitations in this study
and aim to address them in future work. First, \bench focuses on a limited set of 3GPP specifications in Release 17. 
While this design scope enables controlled and reproducible evaluation, it does not cover the full breadth of the 3GPP corpus or reflect the ongoing evolution of specifications across releases. Moreover, we have not evaluated \bench on standards
published by other organizations, such as IEEE, ISO, or IETF.
Second, although \bench incorporates real-world vulnerabilities reported in prior work to construct security-focused tasks, its coverage is inherently limited to publicly documented and curated cases. 
Third, for tasks that require explicit evidence sets, \bench prioritizes verifiability by enforcing strict matching against clause, table, and figure identifiers, and, in advanced settings, by requiring complete evidence that jointly supports a conclusion. While this strict evaluation aligns with the goal of trustworthy specification reasoning, it may under-credit partially correct evidence or alternative but valid supporting sets. 

In future work, we plan to mitigate these limitations in two directions. First, we will expand specification coverage to Release 18 and beyond by leveraging the proposed semi-automated pipeline, with expert verification focused on newly introduced or modified clauses; the benchmark will be versioned and maintained alongside 3GPP releases. We will also explore extending its coverage to standards published
by IEEE, ISO, and IETF. Second, we aim to broaden the benchmark's scope by incorporating additional datasets and task types to further close the gap between specification analysis and real-world deployment security.

%% file: latex/conclusion.tex
\section{Conclusion}
This work addresses a key gap in cellular security research: the lack of a reproducible benchmark and a unified methodology for evaluating 3GPP specification understanding and standards-driven security reasoning. We establish \bench, a staged benchmark spanning clause-local comprehension, evidence-grounded answering, and cross-clause, security-relevant reasoning, built from 3GPP specifications and complemented with carefully verified and corrected subsets from prior datasets. We further introduce \ari, a unified framework that supports domain adaptation to 3GPP specifications, retrieval of multimodal evidence with verifiable citations, and integration reasoning for security analysis.

%% file: latex/appendix.tex
\newpage
\appendix

\section*{Ethical Statement}
Our work studies automatic security reasoning over 3GPP specifications and introduces a benchmark related to procedures and security issues. Our framework and benchmark aim to improve defensive monitoring of standards and implementations. Our datasets are derived from 3GPP technical specifications and from publicly reported vulnerabilities in academic and industry works. The dataset does not involve human subjects or personal data. We will release it under the MIT license, requiring users to conform to its terms. We encourage the use of our benchmark for research purposes, and the authors disclaim liability of any misuse.

\section{Glossary of Cellular Network Terminology}
\label{app:glossary}

This glossary defines the specialized cellular network terminology used in the main text, figures, and appendices.

\scriptsize
\renewcommand{\arraystretch}{1.12}

\begin{longtable}{p{0.28\linewidth} p{0.66\linewidth}}
\caption{Glossary of cellular network terminology used in this paper.}
\label{tab:glossary}\\
\toprule
\textbf{Term} & \textbf{Definition} \\
\midrule
\endfirsthead

\toprule
\textbf{Term} & \textbf{Definition} \\
\midrule
\endhead

\bottomrule
\endfoot

\multicolumn{2}{l}{\textbf{Standards and network architecture}}\\
\midrule

3GPP &
The 3rd Generation Partnership Project, which develops technical specifications for cellular systems, including 4G and 5G networks. \\

Technical Specification (TS) &
A 3GPP document that defines network architectures, protocols, procedures, messages, and implementation requirements. \\

Release &
A coordinated version of 3GPP specifications containing a defined set of cellular network features and updates. \\

User Equipment (UE) &
A device that connects to a cellular network, such as a smartphone, modem, or other connected device. \\

Base Station (BS) &
A network node that provides radio connectivity to UEs. An LTE base station is commonly called an eNodeB, while a 5G NR base station is commonly called a gNodeB. \\

Radio Access Network (RAN) &
The part of a cellular network that provides radio connectivity between UEs and the core network. \\

Core Network &
The part of a cellular network responsible for authentication, mobility management, session management, and data connectivity. \\

Control Plane &
The signaling functions used to establish, configure, manage, and release network connections and services. \\

5G System (5GS) &
The overall 5G network system, including the 5G core network, radio access network, and UEs. \\

Evolved Packet System (EPS) &
The system architecture underlying LTE networks, including the E-UTRAN radio access network and the Evolved Packet Core. \\

Long-Term Evolution (LTE) &
A cellular radio access technology commonly associated with 4G networks. \\

Radio Access Technology (RAT) &
The radio technology used by a UE to access a cellular network, such as LTE or 5G NR. An inter-RAT procedure involves movement between different radio technologies. \\

Public Land Mobile Network (PLMN) &
A cellular network identified by a combination of a mobile country code and a mobile network code. \\

\midrule
\multicolumn{2}{l}{\textbf{Protocols and network functions}}\\
\midrule

Access Stratum (AS) &
The protocol layers and signaling between a UE and the radio access network. RRC belongs to the access stratum. \\

Non-Access Stratum (NAS) &
Control-plane signaling between a UE and the core network. NAS supports registration, authentication, mobility management, and session management. \\

NAS Signaling &
The exchange of NAS messages between a UE and the core network. \\

Radio Resource Control (RRC) &
A control-plane protocol between a UE and the radio access network that manages radio connections, configurations, state transitions, and access-stratum security. \\

RRC Signaling &
The exchange of RRC messages between a UE and the radio access network to establish, configure, suspend, resume, or release a radio connection. \\

5G Mobility Management (5GMM) &
The 5G NAS sublayer responsible for registration, authentication, mobility management, service requests, and deregistration. \\

EPS Mobility Management (EMM) &
The LTE NAS sublayer responsible for attachment, authentication, mobility management, tracking-area updates, and detachment. \\

EPS Session Management (ESM) &
The LTE NAS sublayer responsible for establishing, modifying, and releasing packet-data connectivity. \\

Access and Mobility Management Function (AMF) &
A 5G core network function that terminates NAS signaling and manages UE registration, authentication, reachability, and mobility. \\

Session Management Function (SMF) &
A 5G core network function that manages the establishment, modification, and release of data sessions. \\

Mobility Management Entity (MME) &
An LTE core network function that terminates NAS signaling and manages UE attachment, authentication, and mobility. \\

IP Connectivity Access Network (IP-CAN) &
An access network and its associated bearer resources that provide IP connectivity to a UE. \\

Wideband S1 (WB-S1) Mode &
An LTE/EPS operating mode using wideband E-UTRA access and the S1 interface. The specification defines separate NAS timer behavior for some WB-S1 enhanced-coverage configurations. \\

\midrule
\multicolumn{2}{l}{\textbf{Messages, procedures, and identifiers}}\\
\midrule

Signaling Message &
A protocol message exchanged between network entities to initiate, control, or complete a cellular procedure. \\

Information Element (IE) &
A structured field within a signaling message that carries information such as an identifier, state, cause value, timer, or capability. \\

Registration &
A 5GMM procedure through which a UE establishes or updates its mobility-management context with the 5G core network. \\

Deregistration &
A 5GMM procedure through which the UE or network releases the UE's mobility-management context. \\

Tracking Area Update (TAU) &
An LTE EMM procedure used by a UE to report mobility or update its registration information with the network. \\

Routing Area Update (RAU) &
A mobility-management procedure used in earlier packet-switched cellular systems to update a UE's routing area. \\

Security Mode Control &
A procedure used to select and activate integrity-protection and ciphering algorithms between a UE and the network. \\

International Mobile Subscriber Identity (IMSI) &
A permanent identifier associated with a mobile subscriber. Exposure of the IMSI can enable subscriber identification or tracking. \\

Subscription Permanent Identifier (SUPI) &
The permanent subscriber identifier used in 5G. An IMSI can be represented as a type of SUPI. \\

Subscription Concealed Identifier (SUCI) &
A privacy-preserving representation of the SUPI used to conceal the subscriber's permanent identity when it is transmitted over the radio interface. \\

Globally Unique Temporary Identifier (GUTI) &
A temporary identifier assigned by the core network to reduce repeated transmission of a subscriber's permanent identity. Its 5G form is called the 5G-GUTI. \\

Tracking Area Identity (TAI) &
An identifier for a tracking area used by the network to manage UE location and mobility. \\

IMSI Catcher &
A rogue or impersonated base station that attempts to obtain a subscriber's IMSI or track the subscriber. \\

\midrule
\multicolumn{2}{l}{\textbf{Protocol states}}\\
\midrule

5GMM-REGISTERED &
A 5GMM state in which a mobility-management context has been established for the UE. The UE may initiate mobility registration updates and service-request procedures. \\

5GMM-REGISTERED-INITIATED &
A 5GMM state entered after the UE starts a registration procedure and while it is waiting for a response from the network. \\

5GMM-DEREGISTERED-INITIATED &
A 5GMM state entered after the UE requests release of its 5GMM context and while it is waiting for the network's response. \\

5GMM-DEREGISTERED &
A 5GMM state in which no 5GMM context is established and the UE's location is unknown to the network. The UE must initiate registration to establish a new 5GMM context. \\

EMM-REGISTERED &
An LTE EMM state in which the UE is attached to the EPS and an EMM context has been established. \\

EMM-DEREGISTERED &
An LTE EMM state in which the UE is detached from the EPS and no active EMM registration is maintained. \\

EMM-REGISTERED.ATTEMPTING-TO-UPDATE-MM &
An EMM substate entered when a combined attach or combined tracking-area update succeeds for EPS services only and further mobility-management updating remains necessary. \\

RRC\_IDLE &
An RRC state in which the UE has no active RRC connection and performs cell selection, reselection, and paging monitoring. \\

RRC\_INACTIVE &
An RRC state in which the UE's RRC connection is suspended and selected UE context is retained to support faster connection resumption. \\

RRC\_CONNECTED &
An RRC state in which an active RRC connection exists between the UE and the radio access network. \\

\midrule
\multicolumn{2}{l}{\textbf{Security concepts and protocol timers}}\\
\midrule

NAS Security Context &
Security information shared between a UE and the core network, including selected algorithms, cryptographic keys, and NAS message counters. \\

NAS Integrity Protection &
A security mechanism that allows the receiver to authenticate a NAS message and detect unauthorized modification. \\

AS Security &
Security protection applied to access-stratum signaling and user-plane communication between a UE and the radio access network. \\

Ciphering &
Encryption applied to signaling or user data to protect its confidentiality during transmission. \\

T3510 &
A UE-side 5GMM timer used to supervise a registration procedure after transmission of a REGISTRATION REQUEST message. \\

T3521 &
A UE-side 5GMM timer used to supervise UE-initiated deregistration. It is started after a DEREGISTRATION REQUEST is sent when a network response is expected. \\

\midrule
\multicolumn{2}{l}{\textbf{Specification and evaluation terminology}}\\
\midrule

Clause or Subclause &
A numbered section of a 3GPP specification that defines protocol behavior, procedures, requirements, or message content. \\

Normative Requirement &
A specification statement that defines required, recommended, or permitted behavior, commonly expressed using terms such as \emph{shall}, \emph{should}, and \emph{may}. \\

Evidence Grounding &
The association of an answer or security conclusion with the specification clauses, tables, or figures that directly support it. \\

Intra-Clause Reasoning &
Reasoning for which the required information is contained within a single specification clause. \\

Cross-Clause Reasoning &
Reasoning that combines information from multiple clauses, tables, figures, or specifications. \\

Multimodal Evidence &
Supporting information expressed through more than one specification format, such as text, tables, and figures. \\

\end{longtable}

\normalsize

\section{External Datasets}
\label{appendix:coverage_summary}
Table~\ref{tab:coverage_summary} summarizes the external datasets selectively adopted by \bench, along with their task formats and the corresponding capabilities evaluated at each stage. 

\begin{table}[ht]
\caption{Overview of external datasets included in \bench}
\label{tab:coverage_summary}
\centering
\scriptsize
\setlength{\tabcolsep}{3pt}
\renewcommand{\arraystretch}{1.15}

\begin{tabular}{p{0.05\columnwidth}p{0.2\columnwidth}p{0.25\columnwidth}p{0.45\columnwidth}}
\toprule
\textbf{Stage} & \textbf{Benchmark(s)} & \textbf{Task} & \textbf{Capability Evaluated} \\
\midrule
1 &
TeleQnA \newline Tele-LLMs (Tele-Eval) &
Multiple-choice reasoning \newline Open-ended answering &
Specifications comprehension \newline Terminology and normative language understanding \\
\midrule
2 &
TSpec-LLM &
Multiple-choice reasoning &
Specifications reasoning \\
\midrule
3 &
5GSC\newline ConTester \newline CellularLint \newline TeleQuAD-Tabular &
Sentence-level classification \newline Sentence-pair semantic \newline NLI \newline Tabular QA &
Fine-grained semantics (security vs.\ non-security) \newline Semantic relations \newline Semantic relations \newline Structured information understanding over specification tables \\

\bottomrule
\end{tabular}
\end{table}

\section{Examples for each task type}
\label{appendix:task_types}

\begin{figure}[ht]
    \centering
    \begin{subfigure}[b]{0.48\columnwidth}
        \centering
        \includegraphics[width=\linewidth,page=1]{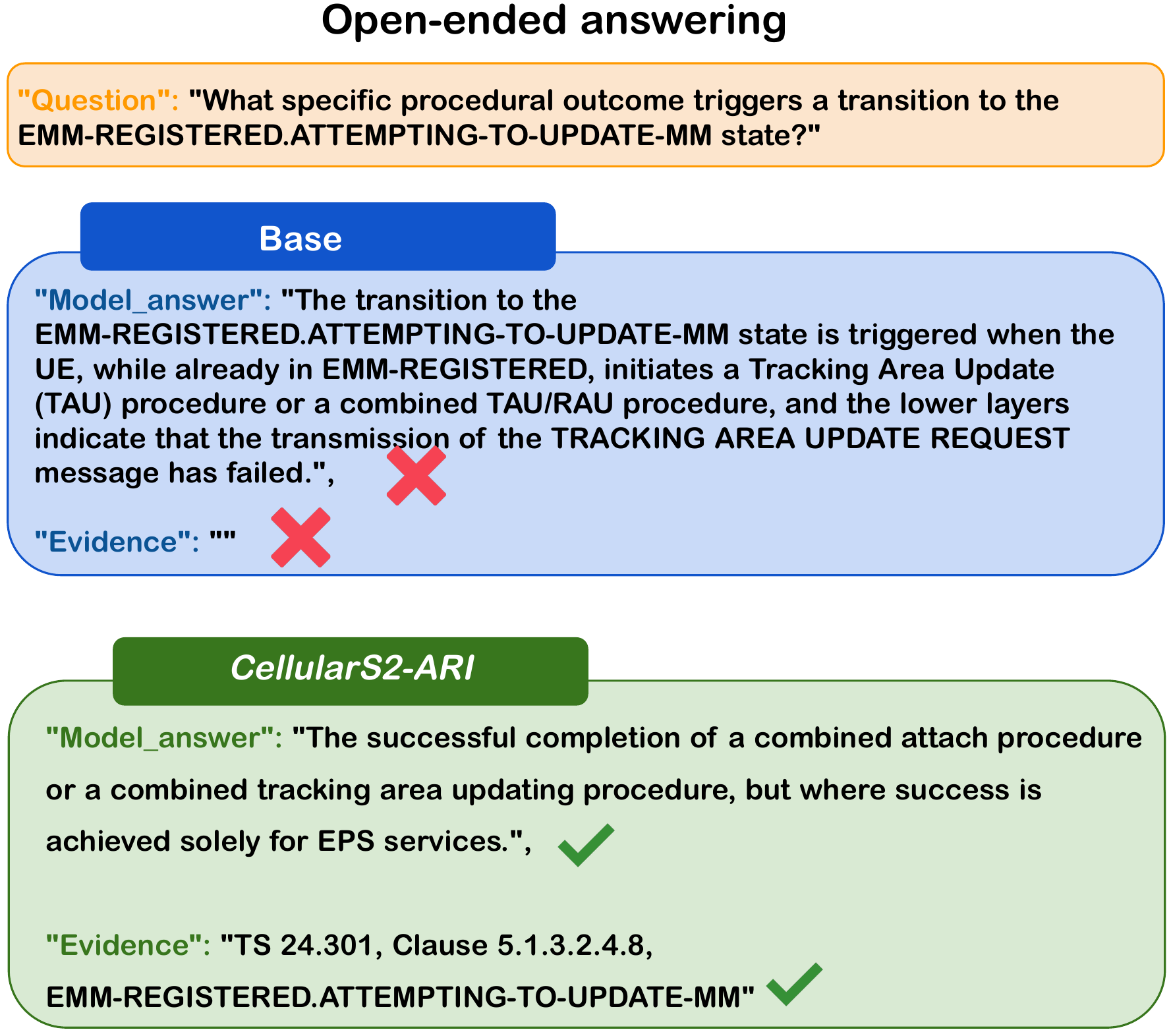}
        \caption{Open-ended answering}
        \label{fig:task_open_ended}
    \end{subfigure}
    \hfill
    \begin{subfigure}[b]{0.48\columnwidth}
        \centering
        \includegraphics[width=\linewidth,page=2]{fig/Task_Types.pdf}
        \caption{Multiple-choice answering}
        \label{fig:task_mcqa}
    \end{subfigure}
    \caption{Task types: open-ended and multiple-choice answering.}
    \label{fig:task_types_combined}
\end{figure}

Here we provide examples for each task type included in \bench, open-ended answering (Figure~\ref{fig:task_open_ended}), multiple-choice answering (Figure~\ref{fig:task_mcqa}), and vulnerability/inconsistency labeling with evidence and explanations (Figures~\ref{fig:task_labeling_ex1}, \ref{fig:task_labeling_ex2}). These figures further present representative cases in which the Base models fail, whereas \ari produces the correct predictions by retrieving and integrating the relevant specification evidence.

\begin{figure}[ht]
    \centering
    \begin{subfigure}[b]{0.48\columnwidth}
        \centering
        \includegraphics[width=\linewidth,page=3]{fig/Task_Types.pdf}
        \caption{Example I}
        \label{fig:task_labeling_ex1}
    \end{subfigure}
    \hfill
    \begin{subfigure}[b]{0.48\columnwidth}
        \centering
        \includegraphics[width=\linewidth,page=4]{fig/Task_Types.pdf}
        \caption{Example II}
        \label{fig:task_labeling_ex2}
    \end{subfigure}
    \caption{Vulnerability/Inconsistency labeling with evidence and explanations.}
    \label{fig:task_labeling_combined}
\end{figure}

\section{Details of Specifications}
\label{appendix:description_of_specification}
\bench intentionally prioritizes security-critical 3GPP
control-plane specifications, particularly the TS~24 series and
TS~38.331. These specifications define foundational cellular procedures,
including registration, authentication, mobility management, security
activation, and radio signaling, where design flaws can directly affect
network availability, privacy, and signaling security. They therefore
provide a practical and high-impact starting point for systematic
specification analysis and benchmark construction.

\noindent $\diamond$ \textbf{TS 24.501 version 17.7.1 (Release 17)~\cite{3GPP_TS_24501}.} It is the 5G NAS (Non-Access Stratum) protocol specification, which governs registration, authentication, security mode control, mobility management, and emergency procedures. It serves as the primary locus of end-to-end signaling protection between the UE and the core network.

\noindent $\diamond$ \textbf{TS 38.331 version 17.0.0 (Release 17)~\cite{3GPP_TS_38331_v1700}.} It is the 5G RRC (Radio Resource Control) protocol, defining radio connection establishment, reconfiguration, and security activation within the access stratum, crucial for understanding where and how radio-layer security is triggered and maintained.

\noindent $\diamond$ \textbf{TS 24.301 version 17.6.0 (Release 17)~\cite{3GPP_TS_24301}.} It is the 4G NAS specification, providing continuity with legacy EPS procedures and exposing downgrade and compatibility behaviors that remain relevant in dual-connectivity and inter-RAT (Radio Access Technology) scenarios.

\section{Question Generation}

Figure \ref{fig:Question_Generation} summarizes the unified prompt template used across all stages of question generation for core control-plane 3GPP specifications.

\label{appendix:question_generation}
\begin{figure}[ht]
    \centering
    \includegraphics[width=0.5\linewidth]{fig/Question_Generation.pdf}
    \caption{\textit{Question Generation}}
    \label{fig:Question_Generation}
\end{figure}

\section{Details of Tasks in \bench}
\label{appendix:stage1-3}

\subsection{Stage~1: Intra-Clause Specification Comprehension}
Tasks in Stage~1 are constructed so that the correct answer can be found within a single specification clause (including all multimodal contexts), without requiring reasoning across multiple clauses, tables, or figures. This stage evaluates the most basic adaptation capability: whether a model can correctly interpret definitions, message formats, parameter descriptions, and normative requirements expressed in 3GPP specifications. The following tasks are included in Stage~1.

\noindent \textbf{Original Tasks.} There are three types of original tasks.  Each task includes 500 instances per specification,
yielding 1,500 instances per task and 4,500 original instances in Stage 1.
\textbf{Extractive QA (EQA)}: It's an open-ended answering task. Models must identify an explicit answer that appears verbatim in a single clause from specifications, measuring literal comprehension of standards; \textbf{Abstractive QA (AQA)}: It's an open-ended answering task. Models must generate a concise, semantically coherent answer that synthesizes information from nearby normative text, measuring high-level understanding beyond span matching; \textbf{Multi-choice QA (MCQA)}: It's a multiple-choice reasoning task. Models need to select the answer that reflects the specification-compliant behavior under a given condition, state, or signaling action. 

\noindent \textbf{Verified and Corrected Tasks.} We incorporate verified and corrected tasks from TeleQnA~\cite{maatouk2023teleqnabenchmarkdatasetassess} and Tele-Eval~\cite{maatouk2025telellmsseriesspecializedlarge}. 
First, we sample multiple-choice questions from TeleQnA, which aggregates questions from heterogeneous sources (e.g., research publications/overviews, lexicons, standards overviews, and standards specifications). Since \bench focuses on specification-grounded understanding, we restrict TeleQnA to the standards-related subsets, including standards overview and standards specifications. We then apply a quality screening to remove ill-posed items and finally sample 500 MCQAs from the remaining pool to form our Stage 1 external MCQA subset. 

Second, we construct a 500-question general QA subset from Tele-Eval. 
Tele-Eval aggregates questions derived from heterogeneous sources, including (i) scientific papers from arXiv, (ii) 3GPP standards, (iii) Wikipedia articles related to telecommunications, and (iv) telecommunications-related websites extracted from Common Crawl dumps. 
Since \bench targets specification-grounded understanding, we restrict our use of Tele-Eval to questions originating from the 3GPP standards subset. 
We then manually inspect the candidate questions to ensure that they are specification-relevant and fall within the scope of \bench. 
After this filtering process, we verify and correct 500 high-quality open-ended answering questions that are fully aligned with our task construction criteria.

The results on these verified and corrected Stage~1 tasks are summarized in Table~\ref{tab:verified_datasets_all}.

\subsection{Stage~2: Evidence-Grounded Answering}

Stage~2 introduces an explicit evidence grounding requirement to enforce the verifiability of model outputs. In this stage, each prediction must be both correct and traceable to the authoritative specification source: the model outputs the answer together with the supporting clause identifier or table/figure identifiers. Evidence grounding prevents answers that are superficially plausible but unverifiable, and therefore measures retrieval-based reasoning in addition to factual correctness. 

\textbf{Original Tasks.} Similar to Stage 1, Stage 2 maintains full coverage across TS 24.501, TS 38.331, and TS 24.301. Across the three specifications, Stage 2 includes 1,500 EQA with Evidence (EQA-E), 1,500 AQA with Evidence (AQA-E), and 1,500 MCQA with Evidence (MCQA-E), aligned with the task definitions in Section~\ref{sec:task_def}. In all cases, \ari must produce a correct answer and an explicit evidence reference, ensuring each prediction is grounded in 3GPP specifications.

\noindent \textbf{Verified and Corrected Tasks.} We include the 3GPP specification portion of TSpec-LLM~\cite{nikbakht2024tspecllmopensourcedatasetllm} to assess robustness under increased question difficulty. 
Because TSpec-LLM does not publicly release its evaluation set, we reproduce its question-generation pipeline using the same GPT-4 API and generate 500 MCQAs grounded in our target specification scope. 
We also considered integrating Telco-DPR~\cite{saraiva2024telcodprhybriddatasetevaluating}; however, manual inspection revealed substantial quality issues, including QA pairs whose answers cannot be located in the referenced 3GPP specifications. We therefore exclude Telco-DPR from the main benchmark to preserve evidence verifiability. These Stage~2 verified and corrected tasks' results are summarized in Table~\ref{tab:verified_datasets_all}.

\subsection{Stage~3: Cross-Clause and Security-Aware Reasoning}
Stage~3 tasks represent the highest-level reasoning requirements in \bench. It requires models to integrate information across multiple specification clauses and perform security-aware interpretation.

\noindent \textbf{Original Tasks.} Stage~3 first includes open-ended answering tasks whose correct solutions require jointly interpreting information across multiple specification clauses and/or combining textual and tabular/figure content.
It includes two task families: Cross-Clause Question Answering (CCQA) and Table-and-Figure Question Answering (TFQA).
CCQAs are constructed only when the correct answer necessarily depends on the combined interpretation of multiple explicitly related clauses. We include both single-unit TFQAs, where the question can be answered using a single table or figure, and cross TFQAs, which require jointly interpreting multiple tables or multiple figures. Under these settings, Stage~3 contains 222 CCQAs and 747 TFQAs. For all questions, models must generate an answer together with an evidence set consisting of clause numbers, when applicable, table identifiers and/or figure labels. 

Stage~3 further introduces two security-analysis tasks: vulnerability/inconsistency labeling, and the evidence and explanation task.  
We first develop a vulnerability set, consolidating from prior works and manually verified~\cite{shaik2015practical, 6550445, kambourakis2011attacks, lee2009detection, leong2014unveiling, kim2019touching, van2015defeating, park2016white, chlosta2019lte, 8958725, michau2016not, hussain20195greasoner, borgaonkar2018new, 8894379, chlosta20215g, hussain2019privacy, al2024hermes, hussain2018lteinspector, 10.1145/3795866.3844474}; non-vulnerable controls are sampled from ordinary normative text.

For the vulnerability/inconsistency labeling task, given a normative sentence, \ari determines whether it encodes a security-relevant vulnerability or a specification inconsistency and, if so, assigns one or more predefined vulnerability or inconsistency categories. 
Based on the vulnerability set, we include 27 vulnerability statements in total: 11 from TS 24.501, 7 from TS 38.331, and 9 from TS 24.301. We provide 50 instances per specification by combining vulnerability cases with non-vulnerable controls. Each output consists of a binary vulnerability label and a multi-label vector over predefined vulnerability categories (e.g., denial-of-service, replay, downgrade, privacy/tracking, spoofing, authentication bypass, etc.). 

Because this is a multi-label task, the 27 vulnerable statements
contain 38 gold category assignments in total. For example, a single
pre-security NAS behavior may simultaneously involve integrity bypass,
identity exposure, and denial of service. Table~\ref{tab:vulnerability_category}
reports per-category performance for the primary DeepSeek-V3.2-Exp +
\ari configuration.

\begin{table}[t]
\caption{Per-category performance of DeepSeek-V3.2-Exp + \ari on
the vulnerability-labeling task.}
\label{tab:vulnerability_category}
\centering
\scriptsize
\begin{tabularx}{\linewidth}{@{}Xrrr@{}}
\toprule
\textbf{Vulnerability category}
& \textbf{\# Gold}
& \textbf{Recall}
& \textbf{Precision} \\
\midrule
Integrity Bypass
& 9 & 88.9\% & 88.9\% \\
Unauthenticated RRC Message
& 10 & 100.0\% & 90.9\% \\
Identity Exposure / Privacy Leakage
& 4 & 75.0\% & 10.7\% \\
Linkability / Tracking
& 2 & 100.0\% & 66.7\% \\
Ciphering Bypass
& 2 & 100.0\% & 100.0\% \\
Ciphering Bypass / Null Encryption
& 2 & 100.0\% & 100.0\% \\
Emergency-Exception Abuse
& 2 & 100.0\% & 100.0\% \\
Privacy / Tracking
& 2 & 100.0\% & 100.0\% \\
Spoofing / False Paging
& 2 & 100.0\% & 100.0\% \\
Unencrypted Broadcast
& 2 & 100.0\% & 100.0\% \\
Fake Reject / Unauthorized Deregistration / DoS
& 1 & 100.0\% & 100.0\% \\
\midrule
\textbf{Total assignments}
& \textbf{38} & -- & -- \\
\bottomrule
\end{tabularx}
\end{table}

Overall, the results show that \ari achieves high recall across
vulnerability categories, correctly identifying 36 of the 38 gold category assignments. It performs particularly well on categories related to ciphering bypass, emergency-exception abuse, spoofing, and unencrypted broadcasts. However, precision varies across categories, with substantial overprediction for Identity Exposure / Privacy Leakage and, to a lesser extent, Linkability / Tracking. This indicates that \ari is effective at detecting security-relevant behaviors but remains less precise when distinguishing semantically overlapping privacy-related categories. Results for categories containing only one or two gold assignments should be interpreted as preliminary because of their limited instances.

The evidence and explanation task, representing the highest reasoning level in \bench, evaluates whether \ari can explain why a vulnerability arises and retrieve the complete set of supporting evidence across multiple clauses/tables/figures. Leveraging the vulnerability set, this task covers 31 vulnerability scenarios in total: 15 from TS 24.501, 9 from TS 38.331, and 7 from TS 24.301. We provide 50 instances per specification by combining vulnerability cases with non-vulnerable controls. For a vulnerable case, \ari outputs (1) a concise explanatory summary and (2) an evidence set consisting of one or more normative clauses (and, when applicable, table identifiers or figure labels); for a non-vulnerable control, the correct output indicates that no vulnerability is present and the evidence set is empty.

\noindent \textbf{Verified and Corrected Tasks.} 
To complement our vulnerability and inconsistency labeling task, Stage~3 incorporates 4 external datasets, each curated through our verification and correction. Results on these verified and corrected Stage~3 tasks are reported in Table~\ref{tab:verified_datasets_all}.

First, 5GSC dataset~\cite{karim2023spec5gdataset5gcellular} contains 2,401 sentences annotated into three classes: Non-Security, Security, and Undefined. Through careful manual inspection, we identify a substantial number of samples whose security labels cannot be inferred from the sentence content itself. For example, the sentence ``IP-CAN Session Termination listed in this Annex'' is labeled as Security in the original dataset, despite merely referring to an annex title without describing any security mechanism, threat, or protection property. To ensure label consistency and semantic clarity, we manually review all samples and remove sentences with insufficient contextual information, including section headers and other low-information fragments. After filtering, we retain 1,454 high-quality sentences.

Second, ConTester~\cite{287360} constructs a dataset by randomly sampling 500 sentences from the 4G NAS specification and forming 400 sentence pairs for semantic classification, reporting an overall accuracy of 97.25\%. However, inspection of the publicly released dataset reveals that it contains only 366 sentence pairs, and that some annotations are inconsistent with standard semantic interpretations. For example, Sample 333 consists of the sentences ``the attach attempt counter is equal to 5'' and ``the UE should set the attach attempt counter to 5'', which are labeled as class 1 (semantically equivalent). In fact, the former describes a system state, whereas the latter specifies an action, indicating a clear semantic distinction. To ensure label consistency and annotation quality, we manually reviewed the dataset and retained 320 sentence pairs.

Third, CellularLint~\cite{298168} provides a sentence-pair dataset derived from 3GPP specifications for Natural Language Inference (NLI). However, its annotation criteria deviate from widely accepted NLI standards. Under standard NLI definitions, entailment requires that the second sentence logically follows from the first, contradiction indicates an explicit conflict within the same context, and neutral denotes plausibility without entailment or contradiction.
We therefore construct a new dataset of 55 sentence pairs sampled from 4G/5G NAS and RRC specifications, with all pairs annotated strictly according to standard NLI. 

Last, we consider TeleQuAD~\cite{telequad2025}, a suite of question-answering datasets for the telecommunications domain, including extractive, generative, retrieval-augmented generation (RAG), and tabular QA tasks. We focus on TeleQuAD-Tabular, which targets question answering over structured specification tables. There are 278 samples. During verification, we remove 3 samples for which the referenced tables cannot be identified in the specifications, and 21 samples whose questions do not allow unambiguous mapping to a specific table. Among the remaining samples, we identify 2 questions with ambiguous answers and 38 instances with incorrect gold answers. We remove one of the ambiguous questions and manually correct the answer of the other, as well as the 38 incorrect gold answers. After filtering and correction, we retain 253 high-quality tabular QA samples.

Table~\ref{tab:verified_datasets_all} summarizes the evaluation results of \ari on external benchmarks across Stage~1--3. 
Note that these benchmarks contain incorrect or inconsistent entries, as discussed previously. All results reported for \ari are therefore computed on our verified and corrected subsets.

Table~\ref{tab:external_verification_examples} presents
representative examples of instances removed, corrected, or re-annotated during expert verification. For TeleQnA and Tele-Eval, we retain only the
specification-grounded subsets aligned with the scope of \bench.
TSpec-LLM is reproduced because its original evaluation set is
unavailable, whereas Telco-DPR is excluded because its answers are unverifiable and contain hallucinated evidence. 

\begin{table}[t]
\caption{Representative verification and correction examples from
external datasets.}
\label{tab:external_verification_examples}
\centering
\scriptsize
\begin{tabularx}{\linewidth}{
    @{}p{1.15cm}p{0.45cm}X p{1.65cm}X@{}}
\toprule
\textbf{Dataset}
& \textbf{ID}
& \textbf{Original item / annotation}
& \textbf{Decision / revised annotation}
& \textbf{Reason} \\
\midrule

5GSC
& 9
& ``IP-CAN Session Termination listed in this Annex.'' Original label:
Security.
& Removed
& The sentence refers only to an annex title and does not independently
describe a security mechanism, threat, or protection property. \\

\addlinespace

ConTester
& 333
& ``The attach attempt counter is equal to 5'' and ``The UE should set
the attach attempt counter to 5.'' Original label:
Semantically Equivalent.
& Corrected to Not Semantically Equivalent
& The first sentence describes a system state, whereas the second
specifies a normative action. \\

\addlinespace

CellularLint
& 1
& A sentence describing when a non-current EPS security context becomes
the current context is paired with a sentence defining the current EPS
security context and explaining eKSI assignment. Original label:
Entailment.
& Re-annotated as Neutral
& The second sentence does not logically follow from the first; it
introduces additional information about native and mapped security
contexts and eKSI assignment. \\

\addlinespace

TeleQuAD-Tabular
& 91
& Question: ``Which feature is associated with the procedure definition
clause 6.2.3 for uplink media streaming?'' Original answer:
``Content publishing.'' Referenced evidence: TS~26.501,
Clause~4.0.1, Table~4.0.1-1.
& Removed
& The referenced table cannot be identified in the corresponding
specification; therefore, the answer and evidence cannot be verified. \\

\bottomrule
\end{tabularx}
\end{table}

\paragraph{Stage 1.}
TeleQnA reports baseline accuracies for the standards/specifications category of 56.97\% (GPT-3.5), 64.78\% (GPT-4), and 69.84\% (GPT-3.5 with standards context). On our verified TeleQnA MCQA subset, \ari achieves 96\% accuracy.

For the open-ended answering subset derived from Tele-Eval, we evaluate answers using a three-level rubric (2: fully correct; 1: partially correct; 0: incorrect), under which \ari attains 487/500 fully correct, 10/500 partially correct, and 3/500 incorrect. 
For comparability with Tele-LLMs (Tele-Eval), which adopts an LLM-as-a-judge binary protocol (Yes/No), we map score=2 to Yes and score=0,1 to No. Under this aligned protocol, \ari achieves a +59.4\% absolute improvement over the base model, compared to the $\sim$25\% improvement reported by Tele-LLMs.

\paragraph{Stage 2.}
Reported results for TSpec-LLM show accuracies of 93\% on easy questions, 65\% on medium questions, and 66\% on hard questions. Under our reproduced evaluation setting, \ari achieves 94\% overall accuracy.

\paragraph{Stage 3.}
On the verified and corrected 5GSC subset, \ari correctly classifies all instances across the three classes, achieving 100\% accuracy, improving over the reported baseline of 79.73\%. 
On the refined ConTester subset, \ari also achieves 100\% accuracy, compared to the reported baseline of 97.25\%. 
On the newly constructed CellularLint dataset aligned with standard definitions, \ari attains 100\% accuracy, compared to the reported baseline of 82.67\%. 
On the curated TeleQuAD-Tabular subset, \ari achieves 78.06\% accuracy. Notably, existing works do not report standardized accuracy metrics for TeleQuAD-Tabular, highlighting the lack of reliable evaluation for table-based reasoning over telecom specifications.

\begin{table*}[t]
\caption{Summary of verified and corrected datasets across Stage~1--3}
\label{tab:verified_datasets_all}
\centering
\scriptsize
\setlength{\tabcolsep}{3pt}
\renewcommand{\arraystretch}{1.15}


\begin{tabular}{
p{0.04\linewidth}
p{0.1\linewidth}
p{0.06\linewidth}
p{0.06\linewidth}
p{0.18\linewidth}
p{0.15\linewidth}
p{0.3\linewidth}}
\toprule
\textbf{Stage} & \textbf{Dataset} & \textbf{Orig.} & \textbf{Kept} & \textbf{Baseline} & \textbf{\ari} & \textbf{Issues} \\
\midrule

1 & TeleQnA & 3000 & 500 &
Std.\ specs: \newline GPT-3.5 56.97\%, \newline GPT-4 64.78\%, \newline GPT-3.5+ctx 69.84\% &
96.00\% &
Restricted to standards overview and specifications; removed non-standard sources; filtered ill-posed items \\

1 & Tele-Eval & 83056 & 500 &
LLM-as-judge (Yes/No); \newline reports +25\% over base &
+59.40\% over base &
Restricted to 3GPP-origin questions; excluded arXiv, Wikipedia, Common Crawl \\

\midrule

2 & TSpec-LLM & 100 & 500 &
Easy 93\%, \newline Medium 65\%, \newline Hard 66\% &
Average 94.00\% &
Reproduced question-generation pipeline due to unavailable evaluation set \\

2 & Telco-DPR & 1741 & 0 &
Acc@10 86.2\% &
--$^\clubsuit$ &
Excluded due to unverifiable answers and hallucinated evidence \\

\midrule

3 & 5GSC & 2,401 & 1,454 &
79.73\% &
100\% &
Insufficient context; label not inferable from sentence alone \\

3 & ConTester & 366 & 320 &
97.25\% &
100\% &
Annotation inconsistency; state vs.\ action conflation \\

3 & CellularLint & 55 & 55 &
82.67\% &
100\% &
Non-standard NLI criteria; re-annotated under standard NLI definitions \\

3 & TeleQuAD-Tabular & 278 & 253 &
--$^*$ &
78.06\% &
Removed samples with unidentifiable tables and ambiguous question-to-table mappings; corrected erroneous and ambiguous gold answers \\

\midrule

\multicolumn{7}{l}{$^\clubsuit$: No result since entire Telco-DPR dataset is excluded from \bench. }\\
\multicolumn{7}{l}{$^*$: No standardized accuracy metrics are reported. } \\
\bottomrule

\end{tabular}
\end{table*}

\section{Annotation and Dataset Construction}
\label{appendix:annotation}

We describe the annotation procedures used to construct and verify the datasets in \bench. Our datasets consist of two groups: (i) original datasets constructed directly from 3GPP specifications, and (ii) datasets derived from existing benchmarks and subsequently verified and corrected by experts.

The original datasets are constructed through a structured four-step process. First, we select in-scope control-plane procedures and message flows guided by known vulnerabilities in prior work, while also including benign procedures to avoid bias toward vulnerability-centric scenarios. Second, we extract normative statements from 3GPP specifications and attach canonical evidence identifiers, including release version, TS number, clause number, and table or figure references; for cases requiring cross-specification reasoning, multiple evidence identifiers are recorded. Third, annotators construct questions and corresponding gold answers using predefined templates (e.g., open-ended QA, multiple-choice QA, and vulnerability or inconsistency reasoning), ensuring that each instance is answerable based on the associated evidence. Finally, each instance is independently audited by \sout{ a separate} annotators to verify answer correctness, evidence validity, and the absence of ambiguity, with any instance failing these checks removed.

For datasets derived from existing benchmarks, we do not adopt them directly. Prior benchmarks often lack consistent expert validation, and we observe that a non-trivial fraction of instances are inaccurate or insufficiently grounded. Therefore, we curate verified subsets through a similar multi-step process. Specifically, we first validate that each instance is within the scope of the target 3GPP specifications, filtering out items originating from heterogeneous sources such as research papers or Wikipedia. We then verify the correctness of gold answers against the specification text, correcting minor issues when possible and excluding unreliable instances. Next, we ensure evidence traceability by checking whether the cited clauses, tables, or figures support the answer; missing or incorrect evidence is corrected or leads to exclusion. All retained instances are then subjected to independent auditing, and only those passing all checks are included.

All annotation and verification processes are conducted by domain experts rather than crowdworkers. In total, five annotators participate, including three responsible for initial construction and verification and two performing independent quality-control audits. All five annotators are domain experts, each with more than five years of research experience in cellular networking
and security and multiple peer-reviewed publications in these areas. Each instance is created by one annotator and subsequently reviewed by two auditors, and an instance is accepted only when both auditors mark it as \textit{Pass}. To quantify annotation reliability, we compute contingency tables $\{a,b,c,d\}$ capturing agreement patterns and report acceptance rates, raw agreement, and Cohen’s $\kappa$.

Across stages, we observe consistently high acceptance rates and strong inter-annotator agreement. For Stage~1, the acceptance rate for original datasets is $4507/4510 = 99.93\%$, with raw agreement of 0.9998 and $\kappa = 0.7999$, while verified external datasets achieve an acceptance rate of $1005/1010 = 99.50\%$ with $\kappa = 0.7490$. For Stage~2, verified external datasets yield an acceptance rate of $1491/1500 = 99.4\%$ with $\kappa = 0.7129$. 
For Stage 3, newly constructed instances achieve an acceptance rate of 1269/1300 = 97.62\%, with raw agreement of 0.9923 and $\kappa = 0.8038$. External candidate instances achieve an acceptance rate of 2981/3000 = 99.37\%, with raw agreement of 0.9967 and $\kappa = 0.6412$. These results demonstrate that the audited datasets are highly reliable and suitable for evaluating evidence-grounded reasoning over 3GPP specifications.

\begin{table}[t]
\caption{Expert auditing statistics}
\label{tab:audit_single_col}
\centering
\scriptsize
\setlength{\tabcolsep}{2pt}
\begin{tabular}{
p{0.1\linewidth}
p{0.1\linewidth}
p{0.25\linewidth}
p{0.15\linewidth}
p{0.15\linewidth}
p{0.15\linewidth}}
\toprule
\textbf{Stage} & \textbf{Group} & \textbf{Audit Table $\{a,b,c,d\}$} & \textbf{Acc.} & \textbf{Raw} & \textbf{$\kappa$} \\
\midrule

\multirow{2}{*}{S1}
& Orig. & $\{4507, 1, 0, 2\}$ & 99.93\% & .9998 & 0.7999 \\
& Ext.  & $\{1005, 1, 1, 3\}$ & 99.50\% & .9980 & 0.7490 \\

\midrule

\multirow{2}{*}{S2}
& Orig. & $\{4507, 1, 0, 2\}$ & 99.93\% & .9998 & 0.7999 \\
& Ext.  & $\{1491, 2, 2, 5\}$ & 99.40\% & .9973 & 0.7129 \\

\midrule

\multirow{2}{*}{S3}
& Orig. & $\{1269, 5, 5, 21\}$ & 97.62\% & .9923 & 0.8038 \\
& Ext.  & $\{2981, 5, 5, 9\}$ & 99.37\% & .9967 & 0.6412 \\

\bottomrule
\end{tabular}
\vspace{-2em}
\end{table}

\section{Extracting multimodal content from 3GPP specifications}
\label{appendix:content_extraction}

\begin{figure}[ht]
    \centering
    \begin{subfigure}[b]{0.48\columnwidth}
        \centering
        \includegraphics[width=\linewidth,page=1]{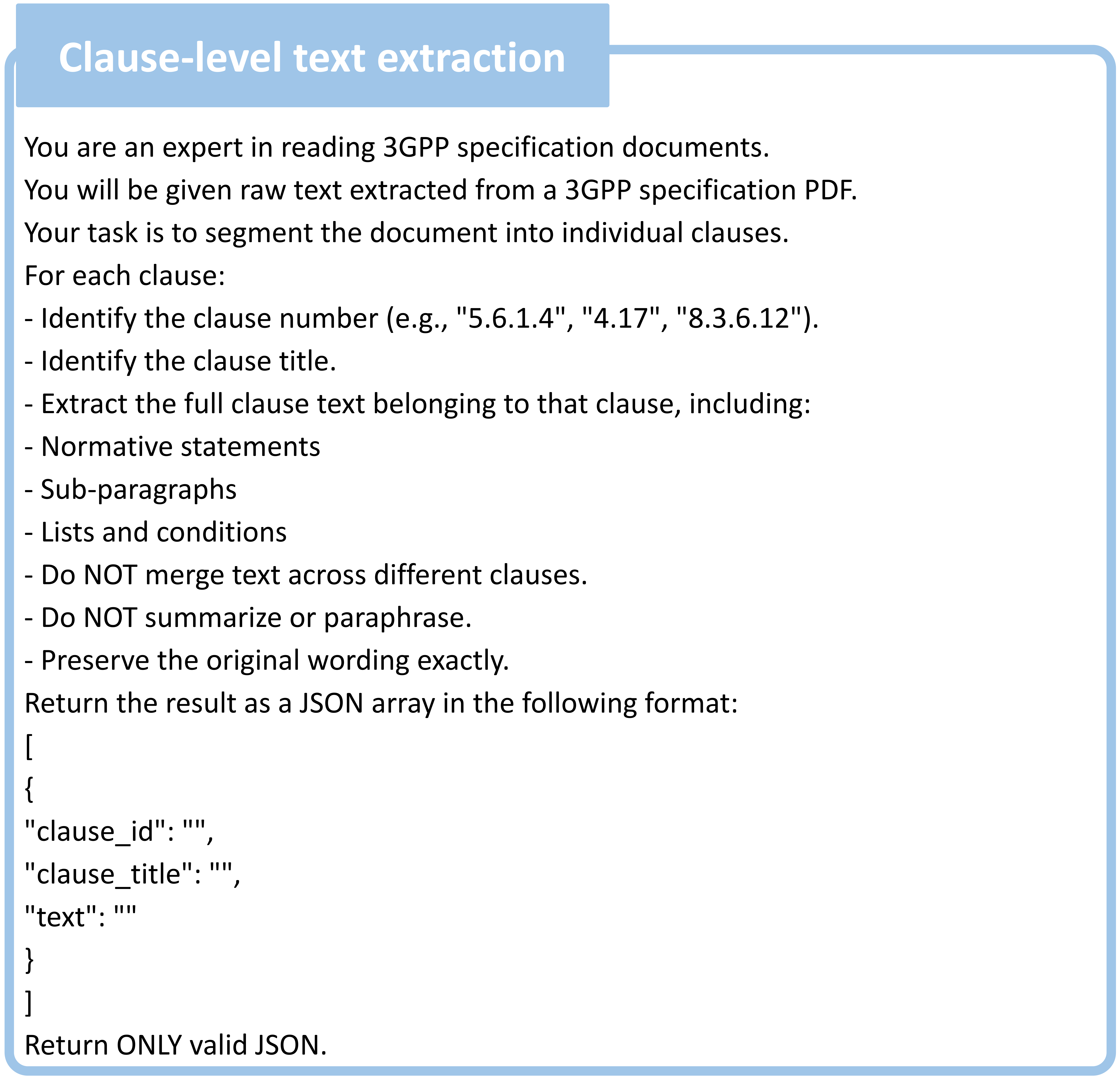}
        \caption{Clause-level text extraction}
        \label{fig:cte}
    \end{subfigure}
    \hfill
    \begin{subfigure}[b]{0.48\columnwidth}
        \centering
        \includegraphics[width=\linewidth,page=2]{fig/benchmark_prompt.pdf}
        \caption{Figure and table identification extraction}
        \label{fig:fti}
    \end{subfigure}
    \caption{Clause-level text extraction and figure, table identification extraction}
    \label{fig:Text_table_figure_extraction}
\end{figure}

Figure~\ref{fig:cte}, \ref{fig:fti}, \ref{fig:figurecontent_extraction}, \ref{fig:table_content_extraction} present the key prompts used to extract and convert texts, tables, and figures from 3GPP specifications into the structured JSON representations. 

\begin{figure}[ht]
    \centering
    \begin{subfigure}[b]{0.48\columnwidth}
        \centering
        \includegraphics[width=\linewidth,page=3]{fig/benchmark_prompt.pdf}
        \caption{Figure content extraction}
        \label{fig:figurecontent_extraction}
    \end{subfigure}
    \hfill
    \begin{subfigure}[b]{0.48\columnwidth}
        \centering
        \includegraphics[width=\linewidth,page=4]{fig/benchmark_prompt.pdf}
        \caption{Table content extraction}
        \label{fig:table_content_extraction}
    \end{subfigure}
    \caption{Figure and table content extraction}
    \label{fig:fig_table_extraction}
\end{figure}

\newpage

\section{Impact of retrieval context size $k$}
\label{appendix:top-k}

To examine the sensitivity of \ari to retrieval context size, we compare its performance under two settings, $k=8$ and $k=6$. All experiments use DeepSeek-V3.2-Exp, with the retrieval weighting
parameter fixed at $\alpha=0.7$ so that only the effect of $k$ is examined. Table~\ref{tab:topk_ablation} summarizes the results on Stage~2 and Stage~3 tasks. All results are reported on the first 50 instances of each task, using the same task definitions and evaluation metrics as in the main experiments. 

We observe that reducing the retrieval context size from $k=8$ to $k=6$ consistently degrades performance across both Stage~2 and Stage~3 tasks, with a more pronounced impact on evidence-intensive tasks. First, tasks that require precise evidence grounding, such as AQA-E and MCQA-E, exhibit the largest performance drops, particularly in evidence accuracy. For example, AQA-E evidence accuracy decreases from 96\% to 72.0\%. This suggests that a smaller retrieval context may omit critical specification clauses needed for correct reasoning and justification. Second, although answer accuracy also decreases, its degradation is generally less severe than that of evidence quality. This indicates that models may still generate plausible answers with limited context, but struggle to support them with correct evidence. Third, Stage~3 tasks show heterogeneous sensitivity to $k$. Complex reasoning tasks such as CCQA exhibit substantial drops in answer accuracy, suggesting that tasks requiring multi-hop reasoning or cross-clause aggregation are particularly sensitive to retrieval coverage. In contrast, tasks focused on explanation or evidence generation show comparatively smaller degradation, indicating that partial context may still support reasonable explanations, although completeness may be affected. Overall, these results show that larger retrieval context sizes improve both answer correctness and evidence grounding, especially for tasks that require precise and comprehensive specification reasoning.

\begin{table}[ht]
\caption{Impact of retrieval context size ($k$) on Stage 2 and 3 tasks}
\label{tab:topk_ablation}
\centering
\scriptsize
\setlength{\tabcolsep}{4pt}
\renewcommand{\arraystretch}{1.15}
\begin{tabular}{p{0.1\columnwidth}p{0.3\columnwidth}p{0.25\columnwidth}p{0.25\columnwidth}}
\toprule
\textbf{Stage} & \textbf{Task} & \textbf{$k=8$} & \textbf{$k=6$} \\
\midrule

2 &
AQA-E &
Score=2: 96\%;  E: 96\% &
Score=2: 88\%; E: 72\% \\

2 &
EQA-E &
Score=2: 96\%; E: 98\% &
Score=2: 86\%; E: 86\% \\

2 &
MCQA-E &
Acc: 100\%; E: 100\% &
Acc: 88\%; E: 86\% \\

\midrule

3 &
CCQA &
Score=2: 94\%; E: 90\% &
Score=2: 74\%; E: 88\% \\

3 &
TFQA &
Score=2: 92\%; E: 96\% &
Score=2: 88\%; E: 92\% \\

3 &
Vulnerability / Inconsistency Labeling &
Bin/Multi: 96\% / 90\% &
Bin/Multi: 88\% / 86\% \\

3 &
Evidence and Explanations &
E: 84\% & E: 80\% \\

\bottomrule
\end{tabular}
\end{table}

\section{The example of taxonomy-based KG}
\label{append:Taxonomy-based KG}
The taxonomy-based KG defines a finite set of entity and relation types that capture the recurring structural, procedural, and security-relevant concepts in 3GPP specifications.
Tables~\ref{tab:entity-types} and~\ref{tab:relation-types} summarize the entity types and relation types used in \textit{SpecReasoning}, respectively.

\begin{table}[ht]
\caption{Entity types defined in \textit{SpecReasoning}}
\label{tab:entity-types}
\centering
\scriptsize
\begin{tabular}{p{0.25\columnwidth}p{0.7\columnwidth}}
\toprule
\textbf{Entity Type} & \textbf{Description} \\
\midrule
Procedure & Representing complete signaling workflows (e.g., Registration, Security Mode Control).\\ 
Message & Denoting individual NAS or RRC signaling messages (e.g., REGISTRATION REQUEST, DEREGISTRATION ACCEPT).\\
Information Element (IE) & Representing message-level fields.\\
Identifier & Covering identifiers such as SUCI or GUTI.\\
Timer & Corresponding to protocol timers (e.g., T3510, T3521).\\
State & Denoting UE or network states (e.g., 5GMM-DEREGISTERED).\\
Condition & Expressing logical or optional constraints (e.g., Presence: O).\\
Property & Capturing higher-level system or security attributes (e.g., NAS Integrity Protection).\\
\bottomrule
\end{tabular}
\end{table}

\begin{table}[ht]
\caption{Relation types defined in \textit{SpecReasoning}}
\label{tab:relation-types}
\centering
\scriptsize
\begin{tabular}{p{0.25\columnwidth}p{0.7\columnwidth}}
\toprule
\textbf{Relation Type} & \textbf{Description} \\
\midrule
\multicolumn{2}{l}{\emph{Core relations} $\mathcal{R}_{core}$} \\
\midrule
has\_step  & Connects a procedure to its constituent signaling steps. \\
contains\_IE & Associates a message with its contained information elements. \\
requires\_state & Execution preconditions (e.g., UE must be in state $s$). \\
starts / stops / resets & Timer operations induced by a step/message. \\
guards\_by & Conditional constraints (e.g., Presence/If-Then). \\
establishes & Creation/activation of contexts (e.g., NAS security context). \\
defined\_in & Links each fact to its source clause/table/figure reference. \\
\midrule
\multicolumn{2}{l}{\emph{Security relations} $\mathcal{R}_{sec}$} \\
\midrule
requires\_protection & An object (message/IE/step) shall be integrity/cipher protected. \\
integrity\_required & Integrity protection is mandated for subsequent handling. \\
may\_be\_unprotected & Object may appear before security is established (exceptions). \\
action\_triggered\_by & A network/UE action is triggered upon receiving a (possibly unauthenticated) message. \\
state\_transition & A message/event causes a state transition
(e.g., 5GMM-REGISTERED $\rightarrow$
5GMM-DEREGISTERED-INITIATED).\\
\bottomrule
\end{tabular}
\end{table}

We illustrate how this taxonomy is instantiated in practice through a concrete example drawn from TS 24.501.
Specifically, we show how \textit{SpecReasoning} transforms a UE-initiated de-registration procedure into a set of typed entities and relations, grounding each extracted fact in its normative source. 

From TS 24.501, Clause 5.5.2.2 (UE-initiated De-registration procedure), \textit{SpecReasoning} analyzes the procedural description and the referenced message definitions. The specification states that the procedure is initiated by the UE sending a DEREGISTRATION REQUEST message, that the De-registration Type Information Element indicates whether the de-registration is due to a switch-off, and that the Access Type specifies whether it applies to 3GPP or non-3GPP access. Based on this passage, \textit{SpecReasoning} extracts a procedure node (UE-initiated De-registration), a message node (DEREGISTRATION REQUEST), information-element nodes (De-registration Type and Access Type), the timer node (T3521), and the relevant state nodes (5GMM-REGISTERED, 5GMM-REGISTERED-INITIATED, and 5GMM-DEREGISTERED-INITIATED). It then identifies relationships indicating that the procedure includes a signaling step involving the transmission of the DEREGISTRATION REQUEST message (has\_step), and that the message contains the De-registration Type IE (contains\_IE). For a de-registration request not due to switch-off, the procedure requires the UE to be in the 5GMM-REGISTERED or 5GMM-REGISTERED-INITIATED state (requires\_state). Sending the DEREGISTRATION REQUEST starts timer T3521 (starts) and causes the UE to enter the 5GMM-DEREGISTERED-INITIATED state (state\_transition). The message may be sent without integrity protection under the conditions specified by the standard (may\_be\_unprotected). 

The resulting KG also captures procedural consequences. The AMF action of sending a DEREGISTRATION ACCEPT message is triggered by reception of the DEREGISTRATION REQUEST message, represented by the action\_triggered\_by relation. Upon receiving the DEREGISTRATION ACCEPT message, the UE stops timer T3521 and completes the transition to the 5GMM-DEREGISTERED state, represented by the stops and state\_transition relations. Each extracted entity and relation is anchored to its normative source through the defined\_in relation and the provenance function $\lambda$, which references TS 24.501, Clause 5.5.2.2.


\section{\ari Implementation Details}
\label{appendix:implementation_details}

\subsection{Training-Free Setting and Leakage Controls}

\ari operates in a fully training-free setting. We do not fine-tune any
of the evaluated LLMs on benchmark questions, answer labels, gold answers, or gold evidence annotations. The retrieval index is constructed exclusively from the processed 3GPP specification corpus and does not contain benchmark questions or annotations. During inference, each evaluated LLM receives only the task input, including the question and candidate options when applicable, together with the specification content retrieved by \ari. It does not receive gold answers, gold labels, or gold evidence.

Gold annotations are used only after answer generation to compute
task-specific evaluation metrics. For open-ended tasks, the LLM judge is provided with the gold answer and, when applicable, gold evidence solely for grading; its scores and justifications are not returned to the evaluated LLM or used to revise its outputs. These controls prevent benchmark-specific annotations from influencing retrieval or answer generation while preserving evidence-grounded evaluation.

\subsection{Module Implementation}

Under this training-free evaluation setting, \ari processes and reasons over the 3GPP specification corpus through its three modules: \textit{SpecFusion}, \textit{SpecAnchor}, and \textit{SpecReasoning}. We next describe the implementation of each module, beginning with \textit{SpecFusion}.

Figure~\ref{fig:chunking_specfusion} illustrates the \textit{SpecFusion} preprocessing steps, which normalize raw 3GPP specification text and segment it into sentence-level and paragraph-level components for downstream processing. Figure~\ref{fig:specfusion} presents the overall \textit{SpecFusion} module, highlighting its role in adapting heterogeneous specification content, including text, tables, and figures, into a unified representation.

\begin{figure}[ht]
    \centering
    \begin{subfigure}[b]{0.48\columnwidth}
        \centering
        \includegraphics[width=\linewidth,page=5]{fig/benchmark_prompt.pdf}
        \caption{Specification text normalization and chunking pipeline (\textit{SpecFusion} preprocessing)}
        \label{fig:chunking_specfusion}
    \end{subfigure}
    \hfill
    \begin{subfigure}[b]{0.48\columnwidth}
        \centering
        \includegraphics[width=\linewidth,page=6]{fig/benchmark_prompt.pdf}
        \caption{\textit{SpecFusion}}
        \label{fig:specfusion}
    \end{subfigure}
    \caption{\textit{SpecFusion} preprocessing and pipeline overview}
    \label{fig:specfusion_combined}
\end{figure}

\clearpage
Figure~\ref{fig:ragchunk_SpecAnchor} shows the construction of clause-aware evidence chunks enriched with structural identifiers, enabling precise attribution to specification clauses, tables, and figures. Figure~\ref{fig:SpecAnchor} depicts the \textit{SpecAnchor} module, which performs hybrid retrieval over the unified component space to anchor model outputs to explicit specification evidence for evidence-grounded reasoning.

\begin{figure}[ht]
    \centering
    \begin{subfigure}[b]{0.48\columnwidth}
        \centering
        \includegraphics[width=\linewidth,page=7]{fig/benchmark_prompt.pdf}
        \caption{\textit{SpecAnchor} chunk construction from 3GPP specifications}
        \label{fig:ragchunk_SpecAnchor}
    \end{subfigure}
    \hfill
    \begin{subfigure}[b]{0.48\columnwidth}
        \centering
        \includegraphics[width=\linewidth,page=8]{fig/benchmark_prompt.pdf}
        \caption{\textit{SpecAnchor}}
        \label{fig:SpecAnchor}
    \end{subfigure}
    \caption{\textit{SpecAnchor} chunk construction and pipeline overview}
    \label{fig:SpecAnchor_combined}
\end{figure}

\clearpage
Figure~\ref{fig:kg_specreasoning} illustrates the construction of the taxonomy-based knowledge graph from retrieved multimodal evidence, capturing entities and their structural and security relationships. Figure~\ref{fig:specreasoning} provides an overview of the \textit{SpecReasoning} module, which integrates these structured representations to support cross-clause reasoning and security analysis.

\begin{figure}[ht]
    \centering
    \begin{subfigure}[b]{0.48\columnwidth}
        \centering
        \includegraphics[width=\linewidth,page=9]{fig/benchmark_prompt.pdf}
        \caption{\textit{SpecReasoning} taxonomy-based knowledge graph construction}
        \label{fig:kg_specreasoning}
    \end{subfigure}
    \hfill
    \begin{subfigure}[b]{0.48\columnwidth}
        \centering
        \includegraphics[width=\linewidth,page=10]{fig/benchmark_prompt.pdf}
        \caption{\textit{SpecReasoning}}
        \label{fig:specreasoning}
    \end{subfigure}
    \caption{Overview of \textit{SpecReasoning}, including knowledge graph construction and reasoning pipeline}
    \label{fig:specreasoning_combined}
\end{figure}

\clearpage

\section{Evidence correctness} 
\label{appendix:evidence_correctness}

Stage~2 introduces explicit evidence grounding, and we report evidence
correctness in addition to answer correctness. Evidence correctness is
measured by whether the predicted evidence set (clause numbers and/or
table/figure identifiers) covers all gold references. For tasks that require
multiple evidence units, we consider the prediction correct only if it
includes the complete gold evidence set.

\section{LLM-as-Judge Reliability}
\label{appendix:human_validation}

\begin{figure}[ht]
  \centering
  \includegraphics[width=0.48\linewidth, page=1]{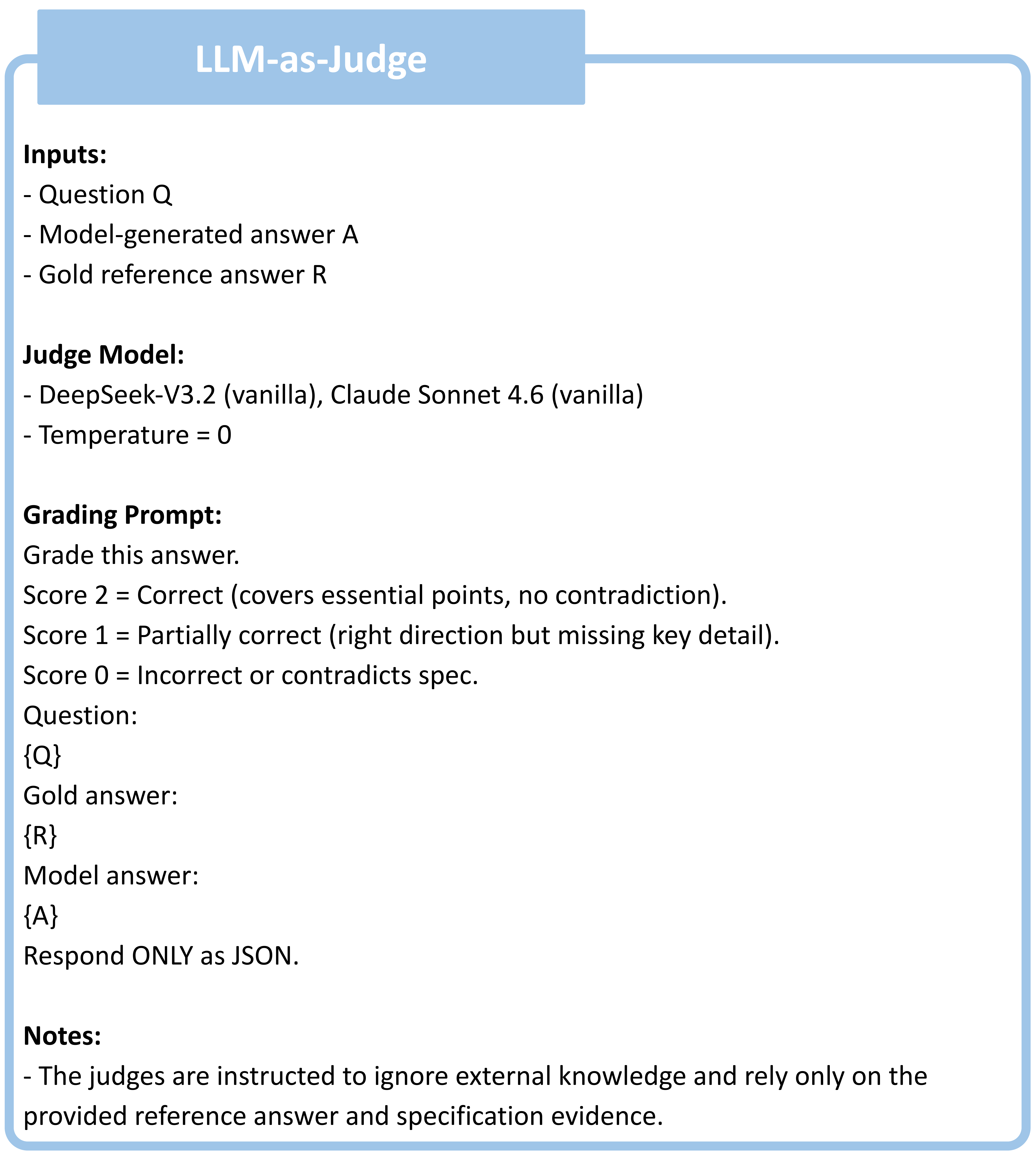}
  \caption{\textit{LLM-as-judge}}
  \label{fig:LLM-as-judge}
\end{figure}

Figure~\ref{fig:LLM-as-judge} presents the prompt template and scoring procedure used by the LLM-based judge for evaluating open-ended question answering tasks. We use automatic judging primarily to support scalable and consistent grading of open-ended outputs. However, using an LLM-as-judge, especially one from the same model family as the evaluated system, may introduce correlated bias and raise reliability concerns. To examine the reliability of LLM-as-judge evaluation, including cases where the judge model comes from the same model family, we conduct a human evaluation study. 

Specifically, we evaluate a representative subset of open-ended instances, sampling 10 questions from each open-ended task. Three annotators independently score model outputs using the same rubric and scoring scale as the automatic judges, based on the gold answer and supporting specification evidence. 

Table~\ref{tab:human_eval} reports (i) inter-annotator agreement among the three human raters using Fleiss' $\kappa$, and (ii) human--judge agreement, measuring how well each automatic judge aligns with the human consensus in terms of accuracy. To further mitigate judge-specific bias, we employ five different judge models: Claude Sonnet
4.6, DeepSeek-V3.2-Exp, Qwen-plus-2025-12-01, Qwen3-30B-A3B-Instruct-2507, and Llama-4-Scout-17B-16E-Instruct. 

We observe several consistent patterns from the human validation results. First, for most tasks, the inter-annotator agreement reaches $\kappa \approx 1.0$, indicating that the evaluation criteria are well-defined and the tasks exhibit low ambiguity. In such cases, annotators consistently assign identical scores, suggesting that both the task formulation and the rubric are sufficiently precise. Second, across all tasks, the relative performance trends across models remain stable under both human evaluation and different automatic judges. In particular, the performance gains of \ari persist across evaluation settings, indicating that the improvements are robust and not tied to a specific judge model. 
Overall, the results show that all judge models exhibit comparable inter-rater reliability and achieve agreement levels close to those of the human annotators. This suggests that our evaluation is not sensitive to the specific choice of judge model, and that different judge models can be used reliably within the proposed framework.

\begin{table}[ht]
\caption{Human evaluation results across all stages}
\label{tab:human_eval}
\centering
\scriptsize
\begin{tabularx}{\linewidth}{
@{}p{0.45cm} p{1.05cm} p{1.75cm} Y Y Y Y Y@{}}
\toprule
\textbf{Stage} & \textbf{Task} & \textbf{Judge}
& \textbf{H1} & \textbf{H2} & \textbf{H3}
& \textbf{AVG} & $\boldsymbol{\kappa}$ \\
\midrule

\multirow{15}{*}{S1}

& \multirow{5}{*}{AQA}
& Claude Sonnet 4.6
& 1.00 & 1.00 & 1.00 & 1.00
& \multirow{5}{*}{1.00} \\
& & DeepSeek-V3.2 & 1.00 & 1.00 & 1.00 & 1.00 & \\
& & Qwen-Plus     & 1.00 & 1.00 & 1.00 & 1.00 & \\
& & Qwen3-30B     & 1.00 & 1.00 & 1.00 & 1.00 & \\
& & Llama-4-17B   & 0.80 & 0.80 & 0.80 & 0.80 & \\

\cmidrule(lr){2-8}

& \multirow{5}{*}{EQA}
& Claude Sonnet 4.6
& 1.00 & 1.00 & 1.00 & 1.00
& \multirow{5}{*}{1.00} \\
& & DeepSeek-V3.2 & 1.00 & 1.00 & 1.00 & 1.00 & \\
& & Qwen-Plus     & 0.80 & 0.80 & 0.80 & 0.80 & \\
& & Qwen3-30B     & 0.80 & 0.80 & 0.80 & 0.80 & \\
& & Llama-4-17B   & 0.80 & 0.80 & 0.80 & 0.80 & \\

\cmidrule(lr){2-8}

& \multirow{5}{*}{Tele-Eval}
& Claude Sonnet 4.6
& 1.00 & 1.00 & 1.00 & 1.00
& \multirow{5}{*}{1.00} \\
& & DeepSeek-V3.2 & 1.00 & 1.00 & 1.00 & 1.00 & \\
& & Qwen-Plus     & 0.60 & 0.60 & 0.60 & 0.60 & \\
& & Qwen3-30B     & 0.60 & 0.60 & 0.60 & 0.60 & \\
& & Llama-4-17B   & 0.20 & 0.20 & 0.20 & 0.20 & \\

\midrule

\multirow{10}{*}{S2}

& \multirow{5}{*}{AQA-E}
& Claude Sonnet 4.6
& 1.00 & 1.00 & 1.00 & 1.00
& \multirow{5}{*}{1.00} \\
& & DeepSeek-V3.2 & 1.00 & 1.00 & 1.00 & 1.00 & \\
& & Qwen-Plus     & 1.00 & 1.00 & 1.00 & 1.00 & \\
& & Qwen3-30B     & 1.00 & 1.00 & 1.00 & 1.00 & \\
& & Llama-4-17B   & 1.00 & 1.00 & 1.00 & 1.00 & \\

\cmidrule(lr){2-8}

& \multirow{5}{*}{EQA-E}
& Claude Sonnet 4.6
& 0.90 & 0.90 & 0.90 & 0.90
& \multirow{5}{*}{$-0.07$} \\
& & DeepSeek-V3.2 & 1.00 & 1.00 & 0.80 & 0.93 & \\
& & Qwen-Plus     & 1.00 & 1.00 & 0.80 & 0.93 & \\
& & Qwen3-30B     & 1.00 & 1.00 & 0.80 & 0.93 & \\
& & Llama-4-17B   & 1.00 & 1.00 & 0.80 & 0.93 & \\

\midrule

\multirow{10}{*}{S3}

& \multirow{5}{*}{CCQA}
& Claude Sonnet 4.6
& 1.00 & 1.00 & 1.00 & 1.00
& \multirow{5}{*}{1.00} \\
& & DeepSeek-V3.2 & 1.00 & 1.00 & 1.00 & 1.00 & \\
& & Qwen-Plus     & 0.40 & 0.40 & 0.40 & 0.40 & \\
& & Qwen3-30B     & 0.40 & 0.40 & 0.40 & 0.40 & \\
& & Llama-4-17B   & 0.60 & 0.60 & 0.60 & 0.60 & \\

\cmidrule(lr){2-8}

& \multirow{5}{*}{TFQA}
& Claude Sonnet 4.6
& 0.90 & 0.90 & 0.90 & 0.90
& \multirow{5}{*}{0.76} \\
& & DeepSeek-V3.2 & 0.80 & 1.00 & 1.00 & 0.93 & \\
& & Qwen-Plus     & 0.60 & 0.60 & 0.60 & 0.60 & \\
& & Qwen3-30B     & 0.60 & 0.60 & 0.60 & 0.60 & \\
& & Llama-4-17B   & 0.40 & 0.40 & 0.40 & 0.40 & \\

\bottomrule
\end{tabularx}
\end{table}

\clearpage

%% file: checklist.tex
\section*{NeurIPS Paper Checklist}

\begin{enumerate}

\item {\bf Claims}
    \item[] Question: Do the main claims made in the abstract and introduction accurately reflect the paper's contributions and scope?
    \item[] Answer: \answerYes{}.
    \item[] Justification: The paper clearly claims (1) \bench: a staged benchmark for 3GPP specification reasoning and (2) \ari: a unified ARI framework. These are consistently supported throughout the paper, including benchmark construction in Section \ref{section:bench}, framework design in Section \ref{section:framework}, and evaluation in Section \ref{section:evaluation}.

\item {\bf Limitations}
    \item[] Question: Does the paper discuss the limitations of the work performed by the authors?
    \item[] Answer: \answerYes{}.
    \item[] Justification: The paper explicitly discusses several limitations in Section \ref{section:limitation_and_future_work}. 

\item {\bf Theory assumptions and proofs}
    \item[] Question: For each theoretical result, does the paper provide the full set of assumptions and a complete (and correct) proof?
    \item[] Answer: \answerNA{}.
    \item[] Justification: This work is primarily an evaluation and datasets paper, not a theoretical paper with formal proofs.

    \item {\bf Experimental result reproducibility}
    \item[] Question: Does the paper fully disclose all the information needed to reproduce the main experimental results of the paper to the extent that it affects the main claims and/or conclusions of the paper (regardless of whether the code and data are provided or not)?
    \item[] Answer: \answerYes{}.
    \item[] Justification: All experiments are reproducible and well-structured, detailed in  Sections~\ref{section:bench}, \ref{section:framework}, and \ref{section:evaluation}

\item {\bf Open access to data and code}
    \item[] Question: Does the paper provide open access to the data and code, with sufficient instructions to faithfully reproduce the main experimental results, as described in supplemental material?
    \item[] Answer: \answerYes{}.
    \item[] Justification: The dataset is available on HuggingFace~\cite{2026cellularspecsecbench_dataset}, and the framework implementation is publicly available on GitHub~\cite{CellSpecSecBench}.

\item {\bf Experimental setting/details}
    \item[] Question: Does the paper specify all the training and test details (e.g., data splits, hyperparameters, how they were chosen, type of optimizer) necessary to understand the results?
    \item[] Answer: \answerYes{}.
    \item[] Justification: The paper clearly describes the data sources (3GPP TS 24.301, TS 24.501, and TS 38.331) in Appendix~\ref{appendix:description_of_specification}, the benchmark construction in Section~\ref{section:bench}, the framework design in Section~\ref{section:framework}, and the hyperparameter settings in Appendix~\ref{appendix:top-k}.

\item {\bf Experiment statistical significance}
    \item[] Question: Does the paper report error bars suitably and correctly defined or other appropriate information about the statistical significance of the experiments?
    \item[] Answer: \answerNA{}
    \item[] Justification: The paper does not report statistical significance measures such as error bars, confidence intervals, or hypothesis testing. 

\item {\bf Experiments compute resources}
    \item[] Question: For each experiment, does the paper provide sufficient information on the computer resources (type of compute workers, memory, time of execution) needed to reproduce the experiments?
    \item[] Answer: \answerYes{}.
    \item[] Justification: The paper specifies the compute environment used for experiments, including the use of an H100 GPU on a Jetstream 2 server. It also describes the model configurations and deterministic inference settings. These details provide sufficient information to understand and reproduce the experimental setup.
    
\item {\bf Code of ethics}
    \item[] Question: Does the research conducted in the paper conform, in every respect, with the NeurIPS Code of Ethics \url{https://neurips.cc/public/EthicsGuidelines}?
    \item[] Answer: \answerYes{}.
    \item[] Justification: The research complies with the NeurIPS Code of Ethics. The work is based on publicly available 3GPP specifications and previously disclosed vulnerabilities, does not involve human subjects or personal data, and is intended to support defensive security analysis. The paper also explicitly addresses ethical considerations and responsible use.

\item {\bf Broader impacts}
    \item[] Question: Does the paper discuss both potential positive societal impacts and negative societal impacts of the work performed?
    \item[] Answer: \answerYes{}.
    \item[] Justification: On the positive side, it aims to improve the security analysis of cellular network specifications, which underpin critical infrastructure used by billions of users. 
    
\item {\bf Safeguards}
    \item[] Question: Does the paper describe safeguards that have been put in place for responsible release of data or models that have a high risk for misuse (e.g., pre-trained language models, image generators, or scraped datasets)?
    \item[] Answer: \answerNA{}.
    \item[] Justification: Our work does not have this risk.

\item {\bf Licenses for existing assets}
    \item[] Question: Are the creators or original owners of assets (e.g., code, data, models), used in the paper, properly credited and are the license and terms of use explicitly mentioned and properly respected?
    \item[] Answer: \answerYes{}.
    \item[] Justification: The paper credits the original sources of external datasets and prior work used in constructing the benchmark, and explicitly identifies the integrated benchmarks and their roles in different stages of \bench. For reused assets, we curate only verified and corrected subsets rather than redistributing them verbatim, and we restrict usage to publicly available specifications and publicly reported materials within their intended research context. Our released benchmark will include its own license and usage terms.

\item {\bf New assets}
    \item[] Question: Are new assets introduced in the paper well documented and is the documentation provided alongside the assets?
    \item[] Answer: \answerYes{}.
    \item[] Justification: The paper introduces new assets, including \bench and the \ari framework. These are well documented in the paper, with detailed descriptions of dataset construction, task design, annotation procedures, and evaluation protocols. The dataset and implementation are publicly released with accompanying documentation to support reproducibility.

\item {\bf Crowdsourcing and research with human subjects}
    \item[] Question: For crowdsourcing experiments and research with human subjects, does the paper include the full text of instructions given to participants and screenshots, if applicable, as well as details about compensation (if any)? 
    \item[] Answer: \answerNA{}.
    \item[] Justification: The paper does not involve crowdsourcing or research with human subjects. Dataset construction and validation are performed by domain experts without human-subject experimentation.

\item {\bf Institutional review board (IRB) approvals or equivalent for research with human subjects}
    \item[] Question: Does the paper describe potential risks incurred by study participants, whether such risks were disclosed to the subjects, and whether Institutional Review Board (IRB) approvals (or an equivalent approval/review based on the requirements of your country or institution) were obtained?
    \item[] Answer: \answerNA{}.
    \item[] Justification: The work does not involve human subjects, user studies, or data collection requiring IRB approval. Therefore, IRB review is not applicable.

\item {\bf Declaration of LLM usage}
    \item[] Question: Does the paper describe the usage of LLMs if it is an important, original, or non-standard component of the core methods in this research? Note that if the LLM is used only for writing, editing, or formatting purposes and does \emph{not} impact the core methodology, scientific rigor, or originality of the research, declaration is not required.
    \item[] Answer: \answerYes{}.
    \item[] Justification: The paper clearly describes the use of large language models as a core component of both dataset construction (e.g., question generation) and the \ari framework. The role of LLMs in adaptation, retrieval, and integration is explicitly detailed, making their usage transparent and central to the methodology.

\end{enumerate}

%% file: latex/custom.bib
@misc{maatouk2023teleqnabenchmarkdatasetassess,
      title={TeleQnA: A Benchmark Dataset to Assess Large Language Models Telecommunications Knowledge}, 
      author={Ali Maatouk and Fadhel Ayed and Nicola Piovesan and Antonio De Domenico and Merouane Debbah and Zhi-Quan Luo},
      year={2023},
      eprint={2310.15051},
      archivePrefix={arXiv},
      primaryClass={cs.IT},
      url={https://arxiv.org/abs/2310.15051}, 
}

@misc{nikbakht2024tspecllmopensourcedatasetllm,
      title={TSpec-LLM: An Open-source Dataset for LLM Understanding of 3GPP Specifications}, 
      author={Rasoul Nikbakht and Mohamed Benzaghta and Giovanni Geraci},
      year={2024},
      eprint={2406.01768},
      archivePrefix={arXiv},
      primaryClass={cs.NI},
      url={https://arxiv.org/abs/2406.01768}, 
}

@misc{saraiva2024telcodprhybriddatasetevaluating,
      title={Telco-DPR: A Hybrid Dataset for Evaluating Retrieval Models of 3GPP Technical Specifications}, 
      author={Thaina Saraiva and Marco Sousa and Pedro Vieira and António Rodrigues},
      year={2024},
      eprint={2410.19790},
      archivePrefix={arXiv},
      primaryClass={cs.IR},
      url={https://arxiv.org/abs/2410.19790}, 
}

@misc{maatouk2025telellmsseriesspecializedlarge,
      title={Tele-LLMs: A Series of Specialized Large Language Models for Telecommunications}, 
      author={Ali Maatouk and Kenny Chirino Ampudia and Rex Ying and Leandros Tassiulas},
      year={2025},
      eprint={2409.05314},
      archivePrefix={arXiv},
      primaryClass={cs.IT},
      url={https://arxiv.org/abs/2409.05314}, 
}

@inproceedings {287360,
author = {Yi Chen and Di Tang and Yepeng Yao and Mingming Zha and XiaoFeng Wang and Xiaozhong Liu and Haixu Tang and Baoxu Liu},
title = {Sherlock on Specs: Building {LTE} Conformance Tests through Automated Reasoning},
booktitle = {32nd USENIX Security Symposium (USENIX Security 23)},
year = {2023},
isbn = {978-1-939133-37-3},
address = {Anaheim, CA},
pages = {3529--3545},
url = {https://www.usenix.org/conference/usenixsecurity23/presentation/chen-yi},
publisher = {USENIX Association},
month = aug
}

@inproceedings {298168,
author = {Mirza Masfiqur Rahman and Imtiaz Karim and Elisa Bertino},
title = {{CellularLint}: A Systematic Approach to Identify Inconsistent Behavior in Cellular Network Specifications},
booktitle = {33rd USENIX Security Symposium (USENIX Security 24)},
year = {2024},
isbn = {978-1-939133-44-1},
address = {Philadelphia, PA},
pages = {5215--5232},
url = {https://www.usenix.org/conference/usenixsecurity24/presentation/rahman},
publisher = {USENIX Association},
month = aug
}

@misc{karim2023spec5gdataset5gcellular,
      title={SPEC5G: A Dataset for 5G Cellular Network Protocol Analysis}, 
      author={Imtiaz Karim and Kazi Samin Mubasshir and Mirza Masfiqur Rahman and Elisa Bertino},
      year={2023},
      eprint={2301.09201},
      archivePrefix={arXiv},
      primaryClass={cs.IR},
      url={https://arxiv.org/abs/2301.09201}, 
}

@techreport{3GPP_TS_24301,
  author      = {{3GPP}},
  title       = {{Universal Mobile Telecommunications System (UMTS); LTE; 5G; Non-Access-Stratum (NAS) protocol for Evolved Packet System (EPS); Stage 3}},
  institution = {{3rd Generation Partnership Project}},
  number      = {{TS 24.301 V17.6.0}},
  year        = {2022},
  note        = {Release 17},
  url         = {http://www.3gpp.org/dynareport/24301.htm}
}

@techreport{3GPP_TS_24501,
  author      = {{3GPP}},
  title       = {{5G; Non-Access-Stratum (NAS) protocol for 5G System (5GS); Stage 3}},
  institution = {{3rd Generation Partnership Project}},
  number      = {{TS 24.501 V17.7.1}},
  year        = {2022},
  note        = {Release 17},
  url         = {http://www.3gpp.org/dynareport/24501.htm}
}

@techreport{3GPP_TS_38331_v1700,
  author      = {{3GPP}},
  title       = {{NR; Radio Resource Control (RRC); Protocol specification}},
  institution = {{3rd Generation Partnership Project}},
  number      = {{TS 38.331 V17.0.0}},
  year        = {2022},
  month       = may,
  note        = {Release 17},
  url         = {https://www.3gpp.org/DynaReport/38331.htm}
}

@misc{colle2025telemathbenchmarklargelanguage,
      title={TeleMath: A Benchmark for Large Language Models in Telecom Mathematical Problem Solving}, 
      author={Vincenzo Colle and Mohamed Sana and Nicola Piovesan and Antonio De Domenico and Fadhel Ayed and Merouane Debbah},
      year={2025},
      eprint={2506.10674},
      archivePrefix={arXiv},
      primaryClass={cs.AI},
      url={https://arxiv.org/abs/2506.10674}, 
}

@software{CellSpecSecBench,
  title = {CellularS2-Bench: A Staged, Evidence-Grounded Benchmark for Cellular Network Security Reasoning},
  author = {CellularS2-Bench},
  organization = {Github},
  year = {2026},
  url = {https://github.com/Chelsiehi/CellularS2}
}

@article{godefroid2020fuzzing,
  title={Fuzzing: Hack, art, and science},
  author={Godefroid, Patrice},
  journal={Communications of the ACM},
  volume={63},
  number={2},
  pages={70--76},
  year={2020},
  publisher={ACM New York, NY, USA}
}

@inproceedings{10.1145/3795866.3844474,
author = {Xie, Ke and Zhao, Xingyi and Chen, Min-Yue and Chen, Yu-An and Hu, Yiwen and Saifuzzaman, Munshi and Li, Wen and Yuan, Shuhan and Tu, Guan-Hua and Xie, Tian},
title = {CellSecInspector: Safeguarding Cellular Networks via Automated Security Analysis on Specifications},
year = {2026},
isbn = {9798400725050},
publisher = {Association for Computing Machinery},
address = {New York, NY, USA},
url = {https://doi.org/10.1145/3795866.3844474},
doi = {10.1145/3795866.3844474},
booktitle = {Proceedings of the 32nd Annual International Conference on Mobile Computing and Networking},
pages = {1098–1122},
numpages = {25},
location = {Austin, TX, USA},
series = {MobiCom '26}
}

@article{shaik2015practical,
  title={Practical attacks against privacy and availability in 4G/LTE mobile communication systems},
  author={Shaik, Altaf and Borgaonkar, Ravishankar and Asokan, N and Niemi, Valtteri and Seifert, Jean-Pierre},
  journal={arXiv preprint arXiv:1510.07563},
  year={2015}
}

@INPROCEEDINGS{6550445,
  author={Bassil, Ramzi and Elhajj, Imad H. and Chehab, Ali and Kayssi, Ayman},
  booktitle={2013 27th International Conference on Advanced Information Networking and Applications Workshops}, 
  title={Effects of Signaling Attacks on LTE Networks}, 
  year={2013},
  volume={},
  number={},
  pages={499-504},
  doi={10.1109/WAINA.2013.136}}

@article{kambourakis2011attacks,
  title={DoS attacks exploiting signaling in UMTS and IMS},
  author={Kambourakis, Georgios and Kolias, Constantinos and Gritzalis, Stefanos and Park, Jong Hyuk},
  journal={Computer Communications},
  volume={34},
  number={3},
  pages={226--235},
  year={2011},
  publisher={Elsevier}
}

@article{lee2009detection,
  title={On the detection of signaling DoS attacks on 3G/WiMax wireless networks},
  author={Lee, Patrick PC and Bu, Tian and Woo, Thomas},
  journal={Computer Networks},
  volume={53},
  number={15},
  pages={2601--2616},
  year={2009},
  publisher={Elsevier}
}

@inproceedings{leong2014unveiling,
  title={Unveiling the hidden dangers of public ip addresses in 4g/lte cellular data networks},
  author={Leong, Wai Kay and Kulkarni, Aditya and Xu, Yin and Leong, Ben},
  booktitle={Proceedings of the 15th Workshop on Mobile Computing Systems and Applications},
  pages={1--6},
  year={2014}
}

@inproceedings{kim2019touching,
  title={Touching the untouchables: Dynamic security analysis of the LTE control plane},
  author={Kim, Hongil and Lee, Jiho and Lee, Eunkyu and Kim, Yongdae},
  booktitle={2019 IEEE Symposium on Security and Privacy (SP)},
  pages={1153--1168},
  year={2019},
  organization={IEEE}
}

@inproceedings{van2015defeating,
  title={Defeating IMSI catchers},
  author={Van Den Broek, Fabian and Verdult, Roel and De Ruiter, Joeri},
  booktitle={Proceedings of the 22Nd ACM SIGSAC Conference on Computer and Communications Security},
  pages={340--351},
  year={2015}
}

@inproceedings{park2016white,
  title={White rabbit in mobile: Effect of unsecured clock source in smartphones},
  author={Park, Shinjo and Shaik, Altaf and Borgaonkar, Ravishankar and Seifert, Jean-Pierre},
  booktitle={Proceedings of the 6th Workshop on Security and Privacy in Smartphones and Mobile Devices},
  pages={13--21},
  year={2016}
}

@inproceedings{chlosta2019lte,
  title={LTE security disabled: misconfiguration in commercial networks},
  author={Chlosta, Merlin and Rupprecht, David and Holz, Thorsten and P{\"o}pper, Christina},
  booktitle={Proceedings of the 12th conference on security and privacy in wireless and mobile networks},
  pages={261--266},
  year={2019}
}

@INPROCEEDINGS{8958725,
  author={Yu, Chuan and Chen, Shuhui},
  booktitle={2019 IEEE 38th International Performance Computing and Communications Conference (IPCCC)}, 
  title={On Effects of Mobility Management Signalling Based DoS Attacks Against LTE Terminals}, 
  year={2019},
  volume={},
  number={},
  pages={1-8},
  doi={10.1109/IPCCC47392.2019.8958725}}

@inproceedings{michau2016not,
  title={How to not break LTE crypto},
  author={Michau, Benoit and Devine, Christophe},
  booktitle={ANSSI Symposium sur la s{\'e}curit{\'e} des technologies de l’information et des communications (SSTIC)},
  year={2016}
}

@inproceedings{hussain20195greasoner,
  title={5GReasoner: A property-directed security and privacy analysis framework for 5G cellular network protocol},
  author={Hussain, Syed Rafiul and Echeverria, Mitziu and Karim, Imtiaz and Chowdhury, Omar and Bertino, Elisa},
  booktitle={Proceedings of the 2019 ACM SIGSAC Conference on Computer and Communications Security},
  pages={669--684},
  year={2019}
}

@article{borgaonkar2018new,
  title={New privacy threat on 3G, 4G, and upcoming 5G AKA protocols},
  author={Borgaonkar, Ravishankar and Hirschi, Lucca and Park, Shinjo and Shaik, Altaf},
  journal={Cryptology ePrint Archive},
  year={2018}
}

@ARTICLE{8894379,
  author={Cao, Jin and Ma, Maode and Li, Hui and Ma, Ruhui and Sun, Yunqing and Yu, Pu and Xiong, Lihui},
  journal={IEEE Communications Surveys \& Tutorials}, 
  title={A Survey on Security Aspects for 3GPP 5G Networks}, 
  year={2020},
  volume={22},
  number={1},
  pages={170-195},
  doi={10.1109/COMST.2019.2951818}}

@inproceedings{chlosta20215g,
  title={5G SUCI-Catchers: Still catching them all?},
  author={Chlosta, Merlin and Rupprecht, David and P{\"o}pper, Christina and Holz, Thorsten},
  booktitle={Proceedings of the 14th ACM Conference on Security and Privacy in Wireless and Mobile Networks},
  pages={359--364},
  year={2021}
}

@article{hussain2019privacy,
  title={Privacy attacks to the 4G and 5G cellular paging protocols using side channel information},
  author={Hussain, Syed Rafiul and Echeverria, Mitziu and Chowdhury, Omar and Li, Ninghui and Bertino, Elisa},
  journal={Network and distributed systems security (NDSS) symposium2019},
  year={2019}
}

@inproceedings{hussain2018lteinspector,
  title={LTEInspector: A systematic approach for adversarial testing of 4G LTE},
  author={Hussain, Syed and Chowdhury, Omar and Mehnaz, Shagufta and Bertino, Elisa},
  booktitle={Network and Distributed Systems Security (NDSS) Symposium 2018},
  year={2018}
}

@inproceedings{al2024hermes,
  title={Hermes: Unlocking security analysis of cellular network protocols by synthesizing finite state machines from natural language specifications},
  author={Al Ishtiaq, Abdullah and Das, Sarkar Snigdha Sarathi and Rashid, Syed Md Mukit and Ranjbar, Ali and Tu, Kai and Wu, Tianwei and Song, Zhezheng and Wang, Weixuan and Akon, Mujtahid and Zhang, Rui and others},
  booktitle={33rd USENIX Security Symposium (USENIX Security 24)},
  pages={4445--4462},
  year={2024}
}

@book{10.5555/7916,
author = {Dreyfus, Hubert L. and Dreyfus, Stuart E. and Athanasiou, Tom},
title = {Mind over machine: the power of human intuition and expertise in the era of the computer},
year = {1986},
isbn = {0029080606},
publisher = {The Free Press},
address = {USA}
}

@misc{jetstream2025llminference,
  title        = {Jetstream2 launches large language model (LLM) inference service},
  howpublished = {Online; Indiana University Jetstream2 News \& Events},
  year         = {2025},
  month        = {apr},
  day          = {11},
  note         = {Accessed on 28 August 2025},
  url          = {https://jetstream-cloud.org/news-events/news/4-11-25_llm-inference-service.html}
}

@misc{deepseek_v3_2_exp_2025,
  title        = {Introducing DeepSeek-V3.2-Exp},
  author       = {{DeepSeek API Docs}},
  year         = {2025},
  month        = sep,
  day          = {29},
  howpublished = {\url{https://api-docs.deepseek.com/news/news250929}},
  note         = {Accessed: 2025-09-29},
  url          = {https://api-docs.deepseek.com/news/news250929}
}

@article{10.1145/3745019,
author = {Gomes, Diego and Felix, Eduardo and Aires, Fernando and Vieira, Marco},
title = {Static Code Analysis for IoT Security: A Systematic Literature Review},
year = {2025},
issue_date = {February 2026},
publisher = {Association for Computing Machinery},
address = {New York, NY, USA},
volume = {58},
number = {3},
issn = {0360-0300},
url = {https://doi.org/10.1145/3745019},
doi = {10.1145/3745019},
journal = {ACM Comput. Surv.},
month = sep,
articleno = {65},
numpages = {47}
}

@misc{telequad2025,
  author       = {Fitsum Gebre and Henrik Holm and
                  Maria Gunnarsson and Doumitrou Nimara and
                  Jieqiang Wei and Vincent Huang and
                  Avantika Sharma and H G Ranjani},
  title        = {{TeleQuAD}: A Suite of Question Answering
                  Datasets for the Telecom Domain},
  year         = {2025},
  howpublished = {Ericsson Research, dataset repository},
  url          = {https://github.com/EricssonResearch/TeleQuAD}
}

@misc{qwen_plus_2025,
  title        = {Qwen-Plus-2025-12-01},
  author       = {{Alibaba Cloud}},
  year         = {2025},
  howpublished = {\url{https://modelstudio.console.alibabacloud.com/}}
}

@misc{llama4_scout_17b_2025,
  title        = {Llama-4-Scout-17B-16E-Instruct},
  author       = {{Meta AI}},
  year         = {2025},
  howpublished = {\url{https://huggingface.co/meta-llama/Llama-4-Scout-17B-16E-Instruct}}
}

@misc{qwen3_30b_a3b_2025,
  title        = {Qwen3-30B-A3B-Instruct-2507},
  author       = {{Alibaba Cloud}},
  year         = {2025},
  howpublished = {\url{https://huggingface.co/Qwen/Qwen3-30B-A3B-Instruct-2507}}
}

@misc{2026cellularspecsecbench_dataset,
  author = {CellularS2-Bench},
  title        = {CellularS2-Bench: A Staged, Evidence-Grounded Benchmark for Cellular Network Security Reasoning},
  year         = {2026},
  howpublished = {\url{https://huggingface.co/datasets/CellularSpecSec-Bench/CellularS2-Bench}},
  note         = {Hugging Face dataset}
}

@techreport{openai_gpt56_2026,
  author      = {{OpenAI}},
  title       = {{GPT-5.6 System Card}},
  institution = {OpenAI},
  year        = {2026},
  month       = jul,
  url         = {https://deploymentsafety.openai.com/gpt-5-6},
  note        = {Accessed: 2026}
}

@techreport{anthropic_sonnet5_2026,
  author      = {{Anthropic}},
  title       = {{Claude Sonnet 5 System Card}},
  institution = {Anthropic},
  year        = {2026},
  month       = jun,
  url         = {https://www-cdn.anthropic.com/480e0bb54327b9622282e9c39a83a4f490ed377e/Claude%20Sonnet%205%20System%20Card.pdf},
  note        = {Accessed: 2026}
}

@techreport{anthropic_sonnet46_2026,
  author      = {{Anthropic}},
  title       = {{Claude Sonnet 4.6 System Card}},
  institution = {Anthropic},
  year        = {2026},
  month       = feb,
  url         = {https://www.anthropic.com/claude-sonnet-4-6-system-card},
  note        = {Accessed: 2026}
}
